# THE $N$–REPRESENTABILITY PROBLEM, THE PSEUDO–SPECTRAL DECOMPOSITION OF ANTISYMMETRIC 1–BODY OPERATORS, AND COLLECTIVE BEHAVIOUR.

Hubert Grudziński [1]
Institute of Physics, Nicholas Copernicus University, Toruń, Poland

The pseudo–spectral decomposition of an $N$–particle antisymmetric 1–body positive–semidefinite operator that corresponds to the canonical convex decomposition into the extreme elements of the dual cone of the set of fermion $N$–representable 1–density operators has been derived. An attempt at constucting a mathematical model for collective behaviour of a system of $N$–fermions that originates from the pseudo–spectral decomposition is presented.

(Abbreviated title: The pseudo–spectral decomposition)

## 1. Introduction

This paper deals with two topics which turn out to be mutually intertwined: the generalized spectral (semi–spectral, non–orthogonal) decomposition of a self–adjoint operator [1], and the fermion $N$–representability problem [4].

The generalized spectral decomposition of a self–adjoint operator has an increasing interest in mathematical physics and leads to the notion of a generalized observable in quantum mechanics, of importance in the quantum theory of measurement (the terms: positive operator–valued measure POV, semi–spectral measure, fuzzy, or unsharp observable are also frequently used) [2, 3, 9, 10, 17, 22, 23, 30, 33, 36].

The generalized spectral decompositions appear naturally in the many–body problem of $N$ fermions interacting through 1– and 2– body forces (p–body in general, $1 \leq p < N$). In this paper we derive some generalized spectral decompositions for a fermion $N$–particle 1–body positive–semidefinite operator acting on a finite dimensional Hilbert space, and give a physical interpretation to the probability measures induced by these decompositions. The main results are concerned with one of the generalized spectral decompositions that appear when the fermion $N$–representability problem is analyzed. We shall call this decomposition a pseudo–spectral decomposition. The pseudo–spectral decomposition has been obtained by a new derivation of the extreme elements of the convex dual cone of the set of fermion $N$–representable 1–densiity operators and reveals the existence of a canonical convex decomposition into the extreme elements of any element belonging to that dual cone. This decomposition induces the pseudo–spectral decomposition. The pseudo–spectral decomposition introduces in the state space of an $N$–fermion system a classification of states into two types: 'particle states', and 'hole states', which in turn leads to the construction of an $N$–fermion 1–body operator possessing 'normal' and 'collective' states as its eigenstates. Such an operator might serve as a mathematical model for the approximate description of the collective behaviour of a system of $N$–fermions if properly adjusted

[1] e–mail address: hubertg@phys.uni.torun.pl

to a physical situation, and may be of interest in superconductivity, magnetic phenomena, collective states of nuclei.

## 2. The $N$–representability problem

In the reduced density matrix approach to the $N$–fermion problem [4, 15, 16, 29, 32], the ground state $\psi^N$ of the $N$–particle 2–body Hamiltonian

$$H^N = \sum_{1\leq i<j\leq N} h^2(i,j)$$

is represented by its 2–particle reduced density operator $D^2(\psi^N)$. The set $\mathcal{P}^2_N$ consisting of all fermion reduced 2–density operators is a proper convex subset of $\mathcal{P}^2$, the set of all fermion 2–density operators. The ground state energy of the system of $N$–fermions could be determined variationally by minimizing the 2–particle functional

$$E = \inf_{D^2\in\mathcal{P}^2_N} \begin{pmatrix} N \\ 2 \end{pmatrix} \mathbf{Tr}\ \left(h^2 D^2\right)$$

over the set $\mathcal{P}^2_N$ instead of the $N$–particle functional

$$E = \inf_{D^N\in\mathcal{P}^N} \mathbf{Tr}\ (H^N D^N)$$

over the set $\mathcal{P}^N$ consisting of all fermion $N$–particle density operators. However, the complete characterization of $\mathcal{P}^2_N$ as a proper convex subset of $\mathcal{P}^2$ is not yet known, and this is the $N$–representability problem [4] for 2–density operators. The elements of $\mathcal{P}^2_N$ are therefore called $N$–representable 2–density operators. It has been shown [25] that the knowledge of all exposed points of $\mathcal{P}^2_N$ is sufficient to characterize the closure of $\mathcal{P}^2_N$. The dual characterization of $\mathcal{P}^2_N$ involves a determination of the dual (polar) cone $\tilde{\mathcal{P}}^2_N$. Any element of $\tilde{\mathcal{P}}^2_N$ provides an $N$–representability condition. Those coming from the extreme elements of $\tilde{\mathcal{P}}^2_N$ are the strongest ones. They give the hyperplane characterization of $\tilde{\mathcal{P}}^2_N$ and thus the solution of the $N$–representability problem. Several necessary conditions for $N$–representability have been derived and some of their structural features and mutual interrelations are established [4–8, 11–16, 18–21, 25–28, 31, 37].

If the Hamiltonian describes a system of $N$ fermions with 1–body interactions only, i.e., $H^N = \sum_{i=1}^N h^1(i)$ (fermion $N$–particle 1–body operator), the 1–particle reduced density operator $D^1$ characterizes the ground state energy $E = \mathbf{inf}\ N\mathbf{Tr}\ (h^1 D^1)$, where minimization is performed with respect to the set $\mathcal{P}^1_N$ consisting of all fermion $N$–representable 1–density operators. The $N$–representability problem for $\mathcal{P}^1_N$ is solved [4, 24]. Namely, $\mathcal{P}^1_N$ is the closed convex hull of the set of all density operators of the form $\frac{1}{N}P^1_{1:N}$, where $P^1_{1:N}$ is a projection operator onto an $N$–dimensional subspace of $\mathcal{H}^1$, where $\mathcal{H}^1$ denotes the 1–particle Hilbert space consisting of functions depending on variables of a single (fermion) particle. Equivalently, a 1–density operator $D^1$ belongs to $\mathcal{P}^1_N$ if and only if $I^1 - ND^1 \geq 0$ ($I^1$: identity on $\mathcal{H}^1$; $\mathbf{Tr}\ D^1 = 1$), or: the eigenvalues of $D^1$ are not greater than $\frac{1}{N}$. $\mathcal{P}^1_N$ being a convex set can be also described by its polar (dual) cone $\tilde{\mathcal{P}}^1_N$ consisting of those 1–particle self–adjoint operators $X^1$ for which $\mathbf{Tr}\ (X^1 D^1) \geq 0$, for all $D^1 \in \mathcal{P}^1_N$ . The dual cone $\tilde{\mathcal{P}}^1_N$ in turn, as a convex cone, is characterized by its extreme rays [4, 24].

In the paper $\mathcal{H}^{\wedge N}$ denotes the Hilbert space consisting of totally antisymmetric functions depending on variables of $N$ fermions. $\mathcal{H}^{\wedge N} = \mathcal{H}^1 \wedge \cdots \wedge \mathcal{H}^1$ ($N$ times) $= A^N(\mathcal{H}^1 \otimes \cdots \otimes \mathcal{H}^1)$, where $A^N$ is the $N$–particle antisymmetrizing operator and $\mathcal{H}^1 \otimes \cdots \otimes \mathcal{H}^1$ is the $N$–fold tensor product of $\mathcal{H}^1$. For $\varphi^p \in \mathcal{H}^{\wedge p}$ and $\psi^q \in \mathcal{H}^{\wedge q}$, $\varphi^p \wedge \psi^q = A^N \varphi^p \otimes \psi^q$. If $X^1$ is a 1–particle operator $X^1 \colon \mathcal{H}^1 \to \mathcal{H}^1$, then $X^{\wedge N} = A^N X^1 \otimes \cdots \otimes X^1 A^N$ ($N$ times) denotes the $N$–th Grassmann power of $X^1$, $X^{\wedge 1} = X^1$. By $X^N$ we denote an arbitrary $N$–particle operator acting on $\mathcal{H}^{\wedge N}$. $I^1$ denotes the 1–particle identity operator acting on $\mathcal{H}^1$, while $I^{\wedge N}$ is the $N$–particle identity operator acting on $\mathcal{H}^{\wedge N}$. By an $N$–particle $p$–body operator we mean any operator of the form $X^p \wedge I^{\wedge(N-p)}$, also denoted by $\Gamma_p^N X^p$ [25], i.e., $\Gamma_p^N X^p = X^p \wedge I^{\wedge(N-p)} = A^N X^p \otimes I^{\otimes(N-p)} A^N$. The mapping $\Gamma_p^N$ is called the $(p,N)$– expansion mapping. In particular, if we compress an $N$–particle 1–body operator $\sum_{i=1}^N X^1(i)$ (more precisely $\sum_{i=1}^N I^1 \otimes \cdots \otimes I^1(i-1) \otimes X^1(i) \otimes I^1(i+1) \otimes \cdots \otimes I^1(N)$) to the antisymmetric space $\mathcal{H}^{\wedge N}$ we get:

$$A^N \sum_{i=1}^N X^1(i) A^N = N X^1 \wedge I^{\wedge(N-1)} = N \Gamma_1^N X^1.$$

In this paper we have given a new derivation of the extreme elements of the dual cone $\tilde{\mathcal{P}}_N^1$ which leads to a rather surprising result of the existence of a canonical convex decomposition into the extreme points of any element belonging to $\tilde{\mathcal{P}}_N^1$. This decomposition induces a generalized spectral (semi––spectral, non–orthogonal) decomposition of any $N$–particle antisymmetric 1–body positive–semidefinite operator $N X^1 \wedge I^{\wedge(N-1)}$ which we call pseudo–spectral decomposition, in order to distinguish it from other semi–spectral decompositions. An attempt at the application of the pseudo–spectral decomposition to the construction of mathematical model for the collective behaviour of a system of $N$–fermions is also given.

## 3. Semi–spectral and spectral decomposition of a self–adjoint $N$–particle antisymmetric 1–body positive–semidefinite operator

We begin with the Lemma that is frequently used in this paper.

**Lemma 3.1:** *Let $P^1$ and $E^1$ be two projection operators onto the mutually orthogonal subspaces $P^1\mathcal{H}^1$ and $E^1\mathcal{H}^1$ of 1–particle Hilbert space $\mathcal{H}^1$. Then for a natural number $N$,*

$$(P^1 + E^1)^{\wedge N} = \sum_{j=0}^{N} \binom{N}{j} P^{\wedge j} \wedge E^{\wedge(N-j)},$$

*where by definition $P^{\wedge 0} \wedge E^{\wedge N} \equiv E^{\wedge N}$, $P^{\wedge N} \wedge E^{\wedge 0} \equiv P^{\wedge N}$ , $P^{\wedge 1} = P^1$, and $E^{\wedge 1} \equiv E^1$. The operators $\binom{N}{j} P^{\wedge j} \wedge E^{\wedge(N-j)}$ $(j = 0, 1, \ldots, N)$ are projectors onto mutually orthogonal subspaces of $\mathcal{H}^{\wedge N}$; if either $j >$ **dim**$(P^1\mathcal{H}^1)$ , or $(N-j) >$ **dim**$(E^1\mathcal{H}^1)$, they are equal to the zero operator.*

*In particular, if $P^1$ and $\tilde{P}^1$ are two mutually orthogonal projectors such that $P^1 + \tilde{P}^1 = I^1$, where $I^1$ is the identity operator on $\mathcal{H}^1$, then the identity operator $I^{\wedge N}$ on $\mathcal{H}^{\wedge N}$ has the decomposition*

$$I^{\wedge N} = \sum_{j=0}^{N} \begin{pmatrix} N \\ j \end{pmatrix} P^{\wedge j} \wedge \tilde{P}^{\wedge(N-j)}.$$

*This decomposition corresponds to the following resolution of the $N$–particle Hilbert space $\mathcal{H}^{\wedge N}$ onto the mutually orthogonal subspaces :*

$$\mathcal{H}^{\wedge N} = \bigoplus_{j=0}^{N} (P^1\mathcal{H}^1)^{\wedge j} \wedge (\tilde{P}^1\mathcal{H}^1)^{\wedge(N-j)}.$$

The statements of the Lemma are certainly known. Unfortunately the author has no reference for the proof except his own [20], which doesn't seem to be the shortest one.

In this paper we analyze some decompositions of operators belonging to two mutually interrelated convex cones of operators: $\tilde{\mathcal{P}}^1_N$ and $\Gamma^N_1 \tilde{\mathcal{P}}^1_N$. The convex cone $\tilde{\mathcal{P}}^1_N$ consists of such 1–particle self–adjoint operators $X^1$ whose antisymmetric $N$–particle expansion $\Gamma^N_1 X^1 \equiv X^1 \wedge I^{\wedge(N-1)} \equiv A^N \left(X^1 \otimes I^{\otimes(N-1)}\right) A^N$ is positive–semidefinite. Here $A^N$ denotes the $N$–particle antisymmetrizing operator. The set of all $X^1 \wedge I^{\wedge(N-1)} \geq 0$ ( i.e. the image of $\tilde{\mathcal{P}}^1_N$ under the expansion mapping $\Gamma^N_1$ ) determines $\Gamma^N_1 \tilde{\mathcal{P}}^1_N$, the convex cone of $N$–particle antisymmetric 1–body positive–semidefinite operators which is a sub–cone of all $N$–particle antisymmetric positive–semidefinite operators $\tilde{\mathcal{P}}^N$, $\Gamma^N_1 \tilde{\mathcal{P}}^1_N \subset \tilde{\mathcal{P}}^N$.

We assume in the paper that the underlying 1–particle Hilbert space $\mathcal{H}^1$ is finite dimensional ($\mathbf{dim}\, \mathcal{H}^1 = n$), and that the 1–particle self-adjoint operator $X^1$ has the following spectral decomposition

$$X^1 = \sum_{i=1}^{s} \beta_i P^1_i + \sum_{i=s+1}^{n} \alpha_i P^1_i, \tag{3.1}$$

where $\beta_i < 0$ $(i = 1, \ldots, s)$, $\beta_i \leq \beta_{i+1}$, and $\alpha_i \geq 0$ $(i = s+1, \ldots, n)$, $\alpha_i \leq \alpha_{i+1}$, $P^1_i$ $(i = 1, \ldots, n)$ are 1–**dim** mutually orthogonal projectors ( $P^1_i P^1_j = P^1_i \delta_{ij}$), and $\sum_{i=1}^{n} P^1_i = I^1$ is the resolution of the identity operator on $\mathcal{H}^1$. The operator (3.1) when expanded to the $N$–particle antisymetric space $\mathcal{H}^{\wedge N}$ has the following decomposition

$$\begin{aligned} NX^1 \wedge I^{\wedge(N-1)} &= \sum_{i=1}^{s} \beta_i N P^1_i \wedge I^{\wedge(N-1)} + \sum_{i=s+1}^{n} \alpha_i N P^1_i \wedge I^{\wedge(N-1)} \\ &= \sum_{i=1}^{s} \beta_i N P^1_i \wedge \tilde{P}^{\wedge(N-1)}_i + \sum_{i=s+1}^{N} \alpha_i N P^1_i \wedge \tilde{P}^{\wedge(N-1)}_i \end{aligned} \tag{3.2}$$

since, due to the Lemma 3.1, $NP^1_i \wedge I^{\wedge(N-1)} = NP^1_i \wedge (P^1_i + \tilde{P}^1_i)^{\wedge(N-1)} = NP^1_i \wedge \tilde{P}^{\wedge(N-1)}_i$, where $P^1_i + \tilde{P}^1_i = I^1$. As we are dealing with cones we prefer to take the operator $NX^1 \wedge I^{\wedge(N-1)}$ instead of just $X^1 \wedge I^{\wedge(N-1)}$ for two reasons:

**(i)** $NP_i^1 \wedge \tilde{P}_i^{\wedge(N-1)} (i = 1, \ldots, n)$ are projection operators (while $P_i^1 \wedge P_i^{\wedge(N-1)}$ are not),

**(ii)** $NX^1 \wedge I^{\wedge(N-1)}$ equals to the appearing in physics $N$–particle 1–body operator $\sum_{i=1}^N X(i)$ when it is compressed to the antisymmetric space $\mathcal{H}^{\wedge N}$.

In the occupation number representation (second quantization notation) Eq. (3.2) looks as

$$NX^1 \wedge I^{\wedge(N-1)} = \sum_{i=1}^{s} \beta_i a_i^+ a_i + \sum_{i=s+1}^{n} \alpha_i a_i^+ a_i .$$

From the mathematical point of view expression (3.2) can be treated as a generalized spectral decomposition of the operator $NX^1 \wedge I^{\wedge(N-1)}$ (the term semi–spectral decomposition, or non––orthogonal spectral decomposition is also frequently used [1, 10, 23, 33]). Namely, the family of positive–semidefinite operators $E_i^N \equiv P_i^1 \wedge I^{\wedge(N-1)} = P_i^1 \wedge \tilde{P}_i^{\wedge(N-1)} (i = 1, \ldots, n)$ constitutes a generalized resolution (non–orthogonal resolution) of the identity operator on $\mathcal{H}^{\wedge N}$: $\sum_{i=1}^n E_i^N = \sum_{i=1}^n P_i^1 \wedge I^{\wedge(N-1)} = I^{\wedge N}$, and generates a normalized positive operator valued (POV)[9, 10] measure $E : \mathcal{A} \to \tilde{\mathcal{P}}^N$, $A_i \equiv \{\omega_i\} \mapsto E_i = P_i^1 \wedge \tilde{P}_i^{\wedge(N-1)}$ on the measurable space $(\Omega, \mathcal{A})$, where $\Omega \equiv \{\omega_i\}_{i=1}^n \equiv \{N\beta_i \, (i = 1, \ldots, s),\, N\alpha_i \, (i = s+1, \ldots, n)\}$, while $\mathcal{A}$ is the $\sigma$–algebra of the subsets of $\Omega$. Thus, Eq. (3.2) can be rewritten in the form

$$NX^1 \wedge I^{\wedge(N-1)} = \sum_{i=1}^{s} (N\beta_i)\, P_i^1 \wedge \tilde{P}_i^{\wedge(N-1)} + \sum_{i=s+1}^{n} (N\alpha_i)\, P_i^1 \wedge \tilde{P}_i^{\wedge(N-1)} = \sum_{i=1}^{n} \omega_i E_i^N . \tag{3.3}$$

Notice, that $E_i^N \, (i = 1, \ldots, n)$ are not projectors, and $E(\Omega) = I^{\wedge N}$.

We have found that from the 'operational' point of view it is more convenient to deal with the non–normalized positive operator valued measure P generated by the non–orthogonal projection operators $P_i^N \equiv NP_i^1 \wedge \tilde{P}_i^{\wedge(N-1)} \; (i = 1, \ldots, n)$, $\sum_{i=1}^n P_i^N = NI^{\wedge N}$, and the value space of P equal to $(\Omega, \mathcal{A})$, where $\Omega = \{\beta_i(i = 1, \ldots, s), \alpha_i(i = s+1, \ldots, n)\}$. Then, Eq. (3.2) can be rewritten in the form

$$NX^1 \wedge I^{\wedge(N-1)} = \sum_{i=1}^{s} \beta_i P_i^N + \sum_{i=s+1}^{n} \alpha_i P_i^N , \tag{3.4}$$

and we will call decomposition (3.4) also a semi–spectral (or generalized spectral) decomposition of the operator $NX^1 \wedge I^{\wedge(N-1)}$. Thus, in the case under consideration the non–normalized POV measure P generated by the operators $P_i^N (i = 1, \ldots, n)$ also determines the operator $NX^1 \wedge I^{\wedge(N-1)}$, provided the generalized resolution of the identity is replaced by the less restrictive requirement $\sum_{i=1}^n P_i^N \geq I^{\wedge N}$, i.e., $P(\Omega) \geq I^{\wedge N}$ (instead of $P(\Omega) = I^{\wedge N}$). This is the normalization requirement $P(\Omega) = \sum_i P_i = I$ that forces projectors to be orthogonal when they appear in the generalized resolution of the identity, i.e., $P_i \geq 0$, $\sum_i P_i = I$, and $P_i^2 = P_i \Rightarrow P_i P_j = 0 (i \neq j)$ [23]. Though this POV measure is not normalized it provides a certain probability measure as it will be seen in Section 7, where also another generalized spectral decomposition of the operator $NX^1 \wedge I^{\wedge(N-1)}$ is analyzed. In order to be able to do that, we need first the spectral (orthogonal) decomposition of this operator.

**Theorem 3.1:** *Let the 1–particle self–adjoint operator $X^1$ has the spectral decomposition*

$$X^1 = \sum_{i=1}^{s} \beta_i P_i^1 + \sum_{i=s+1}^{n} \alpha_i P_i^1, \tag{3.5}$$

*where $\beta_i < 0$ $(i = 1, \ldots, s)$, $\beta_i \leq \beta_{i+1}$; $\alpha_i \geq 0$ $(i = s+1, \ldots, n)$ $\alpha_i \leq \alpha_{i+1}$, $n = dim\mathcal{H}^1$; $P_i^1$:1 – dim projector, $P_i^1 P_j^1 = \delta_{ij} P_i^1$. Then, the operator $NX^1 \wedge I^{\wedge(N-1)}$ (the N–particle antisymmetric expansion of $X^1$) possesses the following spectral (orthogonal) decomposition:*

$$\begin{aligned} NX^1 \wedge I^{\wedge(N-1)} \quad &= \quad \sum_{j=0}^{N} \Bigg( \sum_{1 \leq i_1 < \ldots < i_j \leq s} \; \sum_{s+1 \leq i_{j+1} < \ldots < i_N \leq n} (\beta_{i_1} + \cdots \beta_{i_j} + \alpha_{i_{j+1}} + \cdots + \alpha_{i_N}) \\ &\times \quad N! P_{i_1}^1 \wedge \cdots \wedge P_{i_j}^1 \wedge P_{i_{j+1}}^1 \wedge \cdots \wedge P_{i_N}^1 \Bigg). \end{aligned} \tag{3.6}$$

*Here $N! P_{i_1}^1 \wedge \cdots \wedge P_{i_N}^1$ is a projection operator onto the determinantal state $\sqrt{N!}\,(\varphi_{i_1}^1 \wedge \cdots \wedge \varphi_{i_N}^1) \in \mathcal{H}^{\wedge N}$, $\varphi_i^1 \in \mathcal{H}^1$.*

*The operator is positive–semidefinite if and only if its eigenvalues are non–negative, i.e.,*

$$\beta_{i_1} + \cdots + \beta_{i_j} + \alpha_{i_{j+1}} + \cdots + \alpha_{i_N} \geq 0, \tag{3.7}$$

*where $j = 1, \ldots, s$; $1 \leq i_1 < \ldots < i_j \leq s$; $s+1 \leq i_{j+1} < \ldots < i_N \leq n$. Since the eigenvalues of $X^1$ are ordered in the non–decreasing manner* (3.7) *is equivalent to*

$$\beta_1 + \cdots + \beta_s + \alpha_{s+1} + \cdots + \alpha_N \geq 0. \tag{3.8}$$

The theorem is not a new statement. It is placed here because it serves as a tool in further analysis of the set of 1–body antisymmetric positive–semidefinite operators that determine the dual cone $\tilde{\mathcal{P}}_N^1$ to the convex set of $N$–representable 1–density operators.

The proof of the theorem is placed in Appendix 1 to make the acquaintance of the notation used in the paper, and in its form seems to be original.

Some general features there follow from (3.7) and (3.8) which must possess a 1–particle operator $X^1$, Eq. (3.1), in order to its $N$–particle antisymmetric expansion (3.2) be positive–semidefinite.

1. There cannot be more than $N-1$ eigenvalues $\beta_i$, i.e., $s \leq N-1$.
2. If there are $s$ negative $\beta$'s different from zero, then the dimension of the kernel (the nullspace) of $X^1$ certainly cannot be bigger than $N-s-1$, i.e., $X^1$ with negative eigenvalues and belonging to $\tilde{\mathcal{P}}_N^1$ must have the rank (**dim** of the range) large enough. We will get more precise information regarding the **dim** of the **ker** $X^1$ in Section 5 (the Corollary).

## 4. Extreme elements of the convex cone $\tilde{\mathcal{P}}^1_N$

The set $\tilde{\mathcal{P}}^1_N$ consisting of 1–particle self–adjoint operators $X^1$ whose antisymmetric $N$–particle expansion $N\Gamma^N_1 X^1 \equiv NX^1 \wedge I^{\wedge(N-1)}$ is positive–semidefinite ($\Gamma^N_1 X^1 \geq 0$) is a convex cone, and its extreme elements are the extreme rays. An extreme ray is a positive multiple of an arbitrary element belonging to it which will be called the extreme point. The knowledge of all extreme points characterizes $\tilde{\mathcal{P}}^1_N$, and they are known [4, 24]. Nonetheless, we will give another derivation of the extreme elements of the cone $\tilde{\mathcal{P}}^1_N$, which leads, in the next section, to the rather surprising result of the existence of a canonical convex decomposition into the extreme points of any element belonging to $\tilde{\mathcal{P}}^1_N$. In the case where the underlying Hilbert space is finite dimensional (the one considered in this paper) there is a convenient criterion for describing the extreme elements of $\tilde{\mathcal{P}}^1_N$.

We say that $\mathbf{ker}\,\Gamma^N_1 X^1$ is maximal if there is no another operator $\Gamma^N_1 X^1_0 \geq 0$ such that $\mathbf{ker}\,\Gamma^N_1 X^1 \subset \mathbf{ker}\,\Gamma^N_1 X^1_0$

**Lemma 4.1:** *Let the dimension of the 1–particle Hilbert space be finite. Then,* $X^1 \in \tilde{\mathcal{P}}^1_N$ *is extreme if and only if* $\boldsymbol{ker}\,\Gamma^N_1 X^1$ *is maximal.*

We give an outline of the proof of the Lemma, for the sake of completeness, in Appendix 2 following [8, 14]. For more details the reader is referred to the references. In the Lemma, the finite dimension of the underlying Hilbert space ensures that zero is an isolated point of the spectrum.

By means of Lemma 3.1, Lemma 4.1 and Theorem 3.1, the extreme elements of $\tilde{\mathcal{P}}^1_N$ can be found. Namely, we will find the spectral decomposition of all $N\Gamma^N_1 X^1 \equiv NX^1 \wedge I^{\wedge(N-1)} \geq 0$ possessing a maximal kernel, and then the corresponding $X^1$ which are extreme in $\tilde{\mathcal{P}}^1_N$.

First assume that $X^1 = \sum_{i=1}^n \alpha_i P^1_i$ ($\alpha_i \geq 0$). Then $\Gamma^N_1 X^1$ is positive semidefinite, and $X^1 \in \tilde{\mathcal{P}}^1_N$ is a convex combination of 1–**dim** projectors $P^1_i$ belonging to $\tilde{\mathcal{P}}^1_N$ ($\Gamma^N_1 P^1_i \geq 0$) which are extreme in $\tilde{\mathcal{P}}^1$ (the set of all 1–particle positive–semidefinite operators), and therefore also in $\tilde{\mathcal{P}}^1_N \subset \tilde{\mathcal{P}}^1$. Hence, by virtue of Lemma 4.1, to each extreme element $P^1_i \in \tilde{\mathcal{P}}^1_N$ there corresponds a positive–semidefinite operator $NP^1_i \wedge I^{\wedge(N-1)} = NP^1_i \wedge \tilde{P}^{\wedge(N-1)}_i$ with maximal kernel given by the projector $\tilde{P}^{\wedge N}_i$. This follows from the decomposition $I^{\wedge N} = \left(P^1_i + \tilde{P}^1_i\right)^{\wedge N} = \tilde{P}^{\wedge N}_i + NP^1_i \wedge \tilde{P}^{\wedge(N-1)}_i$ ($\tilde{P}^1_i \equiv I^1 - P^1_i$), where Lemma 3.1 has been used. The result we formulate as

**Proposition 4.1:** *Every 1–dimensional projector* $P^1$ *is an extreme element of* $\tilde{\mathcal{P}}^1_N$.

Now assume $X^1 = \sum_{i=1}^s \beta_i P^1_i + \sum_{j=s+1}^n \alpha_j P^1_j$ ($\beta_i < 0$ $\alpha_j \geq 0$). We observe that all the elements of $\mathcal{H}^1$ that correspond to the projectors $P^1_i$ ($i = 1, \ldots, s$) for which $X^1$ has negative eigenvalues $\beta_i$ ($i = 1, \ldots, s$) must participate in the nullspace $\mathbf{ker}\,X^1 \wedge I^{\wedge(N-1)}$. For if $\beta_1 + \cdots + \beta_{s-1} + \alpha_{i_{s+1}} + \alpha_{i_{s+1}} + \cdots + \alpha_{i_N} = 0$, then $\beta_1 + \cdots + \beta_{s-1} + \beta_s + \alpha_{i_{s+1}} + \cdots + \alpha_{i_N} < 0$ which violates the positive-semidefinitness of $X^1 \wedge I^{\wedge(N-1)}$ (according to Theorem 3.1). Suppose $\beta_1 + \cdots + \beta_s + \alpha_{i_{s+1}} + \cdots + \alpha_{i_N} \equiv \beta + \alpha = 0$. It follows that for a given $\beta$, the maximal kernel requirement of $X^1 \wedge I^{\wedge(N-1)}$ needs all $\alpha_i$ ($i = s{+}1, \ldots, n$) to be equal one to each other, $\alpha_i = \frac{\alpha}{N-s}$ ($i = s+1, \ldots, n$). For if the set of non–negative numbers $\alpha_i$ ($i = s+1, \ldots, n$) satisfying $\alpha_{i_1} + \cdots + \alpha_{i_{N-s}} = \alpha$ consists of unequal numbers, then there exists the minimal number $\alpha_0$ and the sum $\alpha_0+\alpha_{i_2}+\cdots+\alpha_{i_{N-s}} < \alpha_{i_1}+\alpha_{i_2}+\cdots+\alpha_{i_{N-s}}$ (with $\alpha_0 < \alpha_{i_1}$) which contradicts the requirement for the maximal kernel that the all sums are equal to $\alpha$. Therefore, for a fixed negative part $X^1_- \equiv \sum_{i=1}^{s}\beta_i P^1_i$ of a 1–particle operator $X^1 = X^1_- + X^1_+$ the positive part $X^1_+$ must take the form $X^1_+ = \sum_{i=s+1}^{n}\frac{\alpha}{N-s}P^1_i$, $\alpha = -\beta = -(\beta_1+\cdots+\beta_s)$ in order that the corresponding $N$–particle operator $NX^1 \wedge I^{\wedge(N-1)}$ be positive–semidefinite, and with the largest kernel. Then, the corresponding spectral decomposition of $\Gamma_1^N X^1$ is as follows :

$$\begin{aligned} NX^1 \wedge I^{\wedge(N-1)} \quad &= \quad \sum_{j=0}^{s}\Bigg( \sum_{1\le i_1<\ldots<i_j\le s}\ \sum_{s+1\le i_{j+1}<\ldots<i_N\le n} \left(\beta_{i_1}+\ldots+\beta_{i_j}+\frac{N-j}{N-s}\alpha\right) \\ &\times \quad N!P^1_{i_1}\wedge\ldots\wedge P^1_{i_j}\wedge P^1_{i_{j+1}}\wedge\ldots\wedge P^1_{i_N}\Bigg) \ge 0 \end{aligned} \tag{4.1}$$

where the term with $j = s$ is equal to zero ($\beta+\alpha = 0$, $\beta = \beta_1+\cdots+\beta_s$). Hence the projection operator onto the nullspace $\mathbf{ker}\, NX^1 \wedge I^{\wedge(N-1)}$ (lower–case $\mathbf{k}$) is equal to

$$\begin{aligned} \mathbf{Ker}\, NX^1 \wedge I^{\wedge(N-1)} \quad &= \sum_{s+1\le i_{s+1}<\ldots<i_N\le n} N!P^1_1\wedge\ldots\wedge P^1_s\wedge P^1_{i_{s+1}}\wedge\ldots\wedge P^1_{i_N} \\ &= \begin{pmatrix} N \\ s \end{pmatrix} P^{\wedge s}_{1:s}\wedge \tilde{P}^{\wedge(N-s)}_{1:s} \end{aligned}$$

(upper–case $\mathbf{K}$), where $P^1_{1:s} \equiv \sum_{i=1}^{s}P^1_i$, and $\tilde{P}^1_{1:s} \equiv I^1 - P^1_{1:s}$. We denote in this paper a nullspace of an operator $B$, say, by $\mathbf{ker}\, B$ (lower–case $\mathbf{k}$), while we denote the projection operator onto this nullspace by $\mathbf{Ker}\, B$ (upper–case $\mathbf{K}$).

We will show that the operator $NX^1 \wedge I^{\wedge(N-1)} \ge 0$ given by (4.1) attains its maximal kernel when $X^1$ possesses only one negative eigenvalue (all positive are equal one to each other), i.e. $X^1 = -\alpha P^1_1 + \frac{\alpha}{N-1}\sum_{i=2}^{n}P^1_i$ ($\alpha > 0$). To see that, we first reduce the number of negative eigenvalues $\beta_i$ to $s-1$ in (4.1). Then we have the following containment :

$$\mathbf{ker}\, NX^1_{1:s}\wedge I^{\wedge(N-1)} \subset \mathbf{ker}\, NX^1_{1:s-1}\wedge I^{\wedge(N-1)}, \tag{4.2}$$

where

$$\mathbf{Ker}\, NX^1_{1:s}\wedge I^{\wedge(N-1)} = \begin{pmatrix} N \\ s \end{pmatrix} P^{\wedge s}_{1:s}\wedge\tilde{P}^{\wedge(N-s)}_{1:s},$$

$$\mathbf{Ker}\, NX^1_{1:s-1}\wedge I^{\wedge(N-1)} = \begin{pmatrix} N \\ s-1 \end{pmatrix} P^{\wedge(s-1)}_{1:s-1}\wedge\tilde{P}^{\wedge(N-s+1)}_{1:s-1},$$

with

$$X^1_{1:s} \equiv \sum_{i=1}^{s}\beta_i P^1_i + \frac{\alpha}{N-s}\sum_{i=s+1}^{n}P^1_i \quad (\sum_{i=1}^{s}\beta_i+\alpha = 0),$$

and

$$X^1_{1:s-1} \equiv \sum_{i=1}^{s-1}\beta_i P^1_i + \frac{\alpha'}{N-s+1}\sum_{i=s}^{n}P^1_i \quad (\sum_{i=1}^{s-1}\beta_i+\alpha' = 0).$$

For

$$\begin{pmatrix} N \\ s \end{pmatrix} P_{1:s}^{\wedge s} \wedge \tilde{P}_{1:s}^{\wedge(N-s)} = \begin{pmatrix} N \\ s \end{pmatrix} \left(P_{1:s-1}^{1} + P_{s}^{1}\right)^{\wedge s} \wedge \tilde{P}_{1:s}^{\wedge(N-s)}$$
$$= \begin{pmatrix} N \\ s \end{pmatrix} sP_{s}^{1} \wedge P_{1:s-1}^{\wedge(s-1)} \wedge \tilde{P}_{1:s}^{\wedge(N-s)},$$

while

$$\begin{pmatrix} N \\ s-1 \end{pmatrix} P_{1:s-1}^{\wedge(s-1)} \wedge \tilde{P}_{1:s-1}^{\wedge(N-s+1)} = \begin{pmatrix} N \\ s-1 \end{pmatrix} P_{1:s-1}^{\wedge(s-1)} \wedge \left(\tilde{P}_{1:s}^{1} + P_{s}^{1}\right)^{\wedge(N-s+1)}$$
$$= \begin{pmatrix} N \\ s-1 \end{pmatrix} P_{1:s-1}^{\wedge(s-1)} \wedge \Big(\tilde{P}_{1:s}^{\wedge(N-s+1)} + (N-s+1)$$
$$\times \; P_{s}^{1} \wedge \tilde{P}_{1:s}^{\wedge(N-s)}\Big)$$
$$= \begin{pmatrix} N \\ s-1 \end{pmatrix} P_{1:s-1}^{\wedge(s-1)} \wedge \tilde{P}_{1:s}^{\wedge(N-s+1)}$$
$$+ \begin{pmatrix} N \\ s \end{pmatrix} sP_{s}^{1} \wedge P_{1:s-1}^{\wedge(s-1)} \wedge \tilde{P}_{1:s}^{\wedge(N-s)}.$$

Therefore,

$$\mathbf{Ker}\, NX_{1:s-1}^{1} \wedge I^{\wedge(N-1)} = \mathbf{Ker}\, NX_{1:s}^{1} \wedge I^{\wedge(N-1)} + \begin{pmatrix} N \\ s-1 \end{pmatrix} P_{1:s-1}^{\wedge(s-1)} \wedge \tilde{P}_{1:s}^{\wedge(N-s+1)}, \tag{4.3}$$

which says that the containment (4.2) is a proper one (the projection operators on the r.h.s of (4.3) are mutually orthogonal). Proceeding in this way we obtain a chain of nullspaces, the maximal element of which is a nullspace described by the projection operator $\mathbf{Ker}\, X_1^1 \wedge I^{\wedge(N-1)} = NP_1^1 \wedge \tilde{P}_1^{\wedge(N-1)}$. Thus the positive–semidefinite operator $NX_1^1 \wedge I^{\wedge(N-1)}$, with $X_1^1 = -\alpha P_1^1 + \frac{\alpha}{N-1}\sum_{i=2}^{n} P_i^1 = \frac{\alpha}{N-1}\left(I^1 - NP_1^1\right)$ $\alpha > 0$, possesses a maximal kernel. Actually all the operators $\left(I^1 - NP_i^1\right) \wedge I^{\wedge(N-1)} \geq 0$ $\left(i = 1, \ldots, n; \sum_{i=1}^{n} P_i^1 = I\right)$ possess a maximal kernel in $\mathcal{H}^{\wedge N}$. For there is no mutual containment of the kernels described by the projection operators :

$$NP_1^1 \wedge \tilde{P}_1^{\wedge(N-1)} = NP_1^1 \wedge \tilde{P}_{1:2}^{\wedge(N-1)} + \begin{pmatrix} N \\ 2 \end{pmatrix} P_{1:2}^{\wedge 2} \wedge \tilde{P}_{1:2}^{\wedge(N-2)}$$

and

$$NP_2^1 \wedge \tilde{P}_2^{\wedge(N-1)} = NP_2^1 \wedge \tilde{P}_{1:2}^{\wedge(N-1)} + \begin{pmatrix} N \\ 2 \end{pmatrix} P_{1:2}^{\wedge 2} \wedge \tilde{P}_{1:2}^{\wedge(N-2)}$$

as all the different projectors on the r.h.s of both expressions are mutually orthogonal. Here $P_1^1$ and $P_2^1$ stand for the arbitrary two 1–**dim** 1–particle mutually orthogonal projectors. Therefore, in virtue of Lemma 4.1, all $I^1 - NP_i^1$ $(i = 1, \ldots, n)$ are extreme in $\tilde{\mathcal{P}}_N^1$ (since $\left(I^1 - NP_i^1\right) \wedge I^{\wedge(N-1)}$ $(i = 1, \ldots, n)$ possesses maximal kernel). If $\sum_{i=1}^{n} E_i^1 = I^1$ $\left(E_i^1 E_j^1 = E_i \delta_{ij}\right)$ is another spectral decomposition of the 1–particle identity operator $I^1$, then there is no mutual containment of the subspaces described by the projection operators $NP_i^1 \wedge \tilde{P}_i^{\wedge(N-1)}$ and $NE_j^1 \wedge \tilde{E}_j^{\wedge(N-1)}$. Hence we arrive at

**Proposition 4.2:** *For any 1–**dim** projector $P^1$, the operator $\frac{1}{N}I^1 - P^1$ is extreme in $\tilde{\mathcal{P}}_N^1$.*

We have taken the element $\frac{1}{N}I^1 - P^1$ instead of $I^1 - NP^1$ as representing the extreme ray of $\tilde{\mathcal{P}}_N^1$ as being more convenient for further considerations.

## 5. The canonical convex decomposition of $X^1 \in \tilde{\mathcal{P}}_N^1$ into the extreme elements

In order to show that the set $\{P^1, \frac{1}{N}I^1 - P^1; \forall\, P^1\}$ given by Propositions 4.1 and 4.2 contains all the extreme elements of $\tilde{\mathcal{P}}_N^1$ we have to prove that any element $X^1 \in \tilde{\mathcal{P}}_N^1$ can be expressed as a convex combination of the elements belonging to that set. Actually because $\tilde{\mathcal{P}}_N^1$ is a convex cone by a 'convex combination' we mean here a linear combination with non–negative coefficients. In general there is no unique decomposition of an element belonging to a convex set into the extreme points, unless it is a simplex. The following theorem shows that the set $\{P^1,\ \frac{1}{N}I^1 - P^1;\ \forall P^1\}$ exhausts indeed all the extreme elements of the cone $\tilde{\mathcal{P}}_N^1$, giving a prescription for a certain convex (non–negative linear ) decomposition of any element of $\tilde{\mathcal{P}}_N^1$ into these extreme points. We will call it the canonical convex decomposition. Perhaps the term 'semi–convex' instead of 'convex' would be more precise to stress that the non–negative linear combination is not normalized here. On the other hand the term 'the canonical non–negative linear decomposition' does not stress enough the fact that we are dealing with a convex cone. Having this in mind we arrive at the main result of this paper:

**Theorem 5.1:** *If an operator $X^1$ with the spectral decomposition*

$$X^1 = \sum_{i=1}^{s} \beta_i P_i^1 + \sum_{j=s+1}^{n} \alpha_j P_j^1, \tag{5.1}$$

*where $\beta_i < 0$ $(i = 1, \ldots, s)$, $\beta_i \leq \beta_{i+1}$, and $\alpha_j \geq 0$ $(j = s+1, \ldots, n;\ n = \mathbf{dim}\,\mathcal{H}^1)$, $\alpha_j \leq \alpha_{j+1}$, $P_i^1$ $(i = 1, \ldots, n)$ are 1–**dim** orthogonal projectors, belongs to $\tilde{\mathcal{P}}_N^1$ (i.e. $X^1 \wedge I^{\wedge(N-1)} \geq 0$), then it possesses the following convex decomposition into the extreme elements of the cone $\tilde{\mathcal{P}}_N^1$:*

$$X^1 = \sum_{i=1}^{s} \gamma_i \left(\frac{1}{N}I^1 - P_i^1\right) + \sum_{j=s+1}^{r} \gamma_j \left(\frac{1}{N}I^1 - P_j^1\right) + \sum_{k=r+1}^{n} \gamma_k P_k^1. \tag{5.2}$$

*Here $r$ is chosen in such a way that*

$$t_N(r) + \alpha_r < 0, \ \ \textit{while}\ t_N(r) + \alpha_{r+1} \geq 0, \tag{5.3}$$

*where*

$$t_N(r) \equiv \frac{1}{N-r}\left(\sum_{i=1}^{s} \beta_i + \sum_{j=s+1}^{r} \alpha_j\right). \tag{5.4}$$

*Under the above conditions the $\gamma$'s are given by*

$$\gamma_i = -\left(t_N(r) + \beta_i\right) > 0 \quad (i = 1, \ldots, s), \tag{5.5}$$

$$\gamma_j = -\left(t_N(r) + \alpha_j\right) > 0 \quad (j = s+1, \ldots, r), \tag{5.6}$$

$$\gamma_k = t_N(r) + \alpha_k \geq 0 \quad (k = r+1, \ldots, n). \tag{5.7}$$

*Proof.* First we observe that the existence of the inequalities (5.3) is guaranteed by the positive–semi-definiteness of $X^1 \wedge I^{\wedge(N-1)} \geq 0$, which according to Theorem 3.1 is equivalent to

$$\sum_{i=1}^{s} \beta_i + \sum_{j=s+1}^{N} \alpha_j \geq 0, \tag{5.8}$$

and from which it follows that r must satisfy$s+1 \leq r \leq N-1$. (In the case: $\sum_{i=1}^{s} \beta_i + (N-s)\alpha_{s+1} \geq 0$, we set $r = s$, and $t_N(r) = t_N(s) = (N-s)^{-1} \sum_{i=1}^{s} \beta_i$). Rewriting (5.3) in the form

$$\sum_{i=1}^{s} \beta_i + \sum_{j=s+1}^{r} \alpha_j + (N-r)\alpha_r < 0,$$

$$\sum_{i=1}^{s} \beta_i + \sum_{j=s+1}^{r} \alpha_j + (N-r)\alpha_{r+1} \geq 0 \qquad (s+1 \leq r \leq N-1), \tag{5.9}$$

and taking into account that all $\beta_i < 0$ $(i = 1, \ldots, s)$ and $\alpha_j \geq 0$ $(j = s+1, \ldots, N, \ldots, n)$ are ordered in an increasing manner, the inequalities (5.9) follow from (5.8) by inspection. For the case $s = N-1$, the second term in the convex decomposition (5.2) equals zero. Secondly, assuming the inequalities (5.3) are satisfied, it is easy to see that all the $\gamma$'s (5.5)–(5.7) are non–negative:

- **(i)** $\gamma_i > 0$ $(i = 1, \ldots, s)$ since both $t_N(r)$ and $\beta_i$ are strictly negative;

- **(ii)** $\gamma_j > 0$ $(j = s+1, \ldots, r)$, since $0 \leq \alpha_j \leq \alpha_r$;

- **(iii)** $\gamma_k \geq 0$ $(k = r+1, \ldots, n)$ follows again from (5.3) because $\alpha_k \geq \alpha_{r+1}$.

Finally, to show that the $\gamma$'s given by (5.5)–(5.7) are determined correctly, we substitute them into (5.2). Taking into account (5.4), the equality $\sum_{i=1}^{n} P_i^1 = I^1$, and performing the required summations we arrive at (5.1) easily. This completes the proof of the Theorem.□

We observe that Theorem 5.1 provides information concerning the possible maximal dimension of the nullspace of an operator $X^1$ belonging to the cone $\tilde{\mathcal{P}}_N^1$.

**Corollary 5.1:** *Under the conditions of Theorem 5.1,* ***dim ker*** $X^1 \leq r - s$*, i.e., only the eigenvalues* $\alpha_j$ $(j = s+1, \ldots, r)$ *could be equal to zero.*

This can be seen from Eqs. (5.6) and (5.7) remembering that $t_N(r) < 0$. Let us also observe that the arbitrary non–negative linear combination $\sum_{i=1}^{n} \tilde{\omega}_i \left(\frac{1}{N} I^1 - P_i^1\right) + \sum_{i=1}^{n} \omega_i P_i^1$, where $\tilde{\omega}_i \geq 0$ $(i = 1, \ldots, n)$, $\omega_i \geq 0$ $(i = 1, \ldots, n)$, $\sum_{i=1}^{n} P_i^1 = I^1$, of the extreme elements of the cone $\tilde{\mathcal{P}}_N^1$ leads to a certain operator $X^1$ belonging to $\tilde{\mathcal{P}}_N^1$ which in turn, due to Theorem 5.1, can be canonically expressed as a non–negative linear combination of no more than $n$ of the extreme elements. Notice that in the canonical convex decomposition the appearance of the extreme element $\frac{1}{N} I^1 - P_i^1$ eliminates the extreme element $P_i^1$ (with the same subscript 'i'), and vice versa.

## 6. The pseudo–spectral decomposition of an $N$–particle antisymmetric 1–body positive–semidefinite operator

According to the definition of the polar (dual) cone $\tilde{\mathcal{P}}_N^1$ any element $X^1$ of $\tilde{\mathcal{P}}_N^1$ when expanded to the $N$–particle antisymmetric space becomes a positive–semidefinite operator, $NX^1 \wedge I^{\wedge(N-1)} \geq 0$. Therefore, to the canonical convex decomposition of $X^1 \in \tilde{\mathcal{P}}_N^1$ into the extreme elements of $\tilde{\mathcal{P}}_N^1$, given by Eq. (5.2), there corresponds a convex decomposition of the positive–semidefinite operator $NX^1 \wedge I^{\wedge(N-1)}$ into simpler positive–semidefinite operators that are $N$–particle antisymmetric expansions of the extreme elements of $\tilde{\mathcal{P}}_N^1$, i.e., $NP^1 \wedge I^{\wedge(N-1)}$, and$N\left(\frac{1}{N}I^1 - P^1\right) \wedge I^{\wedge(N-1)}$, where the $P^1$'s are 1–**dim** projectors. This decomposition will be called the pseudo–spectral decomposition of an $N$–particle antisymmetric 1–body positive–semidefinite operator $NX^1 \wedge I^{\wedge(N-1)}$ to distinguish it from other semi–spectral decompositions of $NX^1 \wedge I^{\wedge(N-1)}$. Thus we arrive at

**Theorem 6.1:** *If $NX^1 \wedge I^{\wedge(N-1)}$ is an $N$–particle antisymmetric 1–body positive–semidefinite operator then it has the following pseudo–spectral decomposition*

$$\begin{aligned} NX^1 \wedge I^{\wedge(N-1)} &= \sum_{i=1}^{r} \gamma_i \left(I^1 - NP_i^1\right) \wedge I^{\wedge(N-1)} + \sum_{k=r+1}^{n} \gamma_k NP_k^1 \wedge I^{\wedge(N-1)} \\ &= \sum_{i=1}^{r} \gamma_i \tilde{P}_i^{\wedge N} + \sum_{k=r+1}^{n} \gamma_k NP_k^1 \wedge \tilde{P}_k^{\wedge(N-1)}, \end{aligned} \tag{6.1}$$

*where the meaning of $X^1$, $r$, $\gamma_i > 0$ $(i = 1, \ldots, r)$, $\gamma_k \geq 0$, $(k = r+1, \ldots, n)$ is given in Theorem 5.1, while $\tilde{P}_i^{\wedge N}$ and $NP_k^1 \wedge \tilde{P}_k^{\wedge(N-1)}$ are projectors acting on $\mathcal{H}^{\wedge N}$.*

In the occupation number representation Eq. (6.1) looks as follows

$$NX^1 \wedge I^{\wedge(N-1)} = \sum_{i=1}^{r} \gamma_i a_i a_i^+ + \sum_{k=r+1}^{n} \gamma_k a_k^+ a_k .$$

We notice that the operator $NP_i^1 \wedge \tilde{P}_i^{\wedge(N-1)}$ is orthogonal only to $\tilde{P}_i^{\wedge N}$ (with the same subscript i), and it holds that $NP_i^1 \wedge \tilde{P}_i^{\wedge(N-1)} + \tilde{P}_i^{\wedge N} = I^{\wedge N}$, where $I^{\wedge N}$ is the identity operator on $\mathcal{H}^{\wedge N}$. All other pairs of projection operators belonging to the set $\left\{NP_i^1 \wedge \tilde{P}_i^{\wedge(N-1)}, \quad \tilde{P}_i^{\wedge N}\right\}_{i=1}^{n}$ are not mutually orthogonal (i.e., they project onto the subspaces of $\mathcal{H}^{\wedge N}$ which are not mutually orthogonal). Therefore, the decomposition (6.1) is a convex (non–negative linear) decomposition of a positive–semidefinite $N$–particle operator $NX^1 \wedge I^{\wedge(N-1)}$ into the non–orthogonal projection operators corresponding to the extreme elements of the polar cone $\tilde{\mathcal{P}}_N^1$. Similarly to formula (3.2) the decomposition (6.1) can be 'arranged' as a semi–spectral (generalized spectral) decomposition. We define a family of self–adjoint positive–semidefinite operators that constitute a generalized resolution of the identity:

$$F_j^N := \frac{N}{n} P_j^1 \wedge I^{\wedge(N-1)} = \frac{1}{n} NP_j^1 \wedge \tilde{P}_j^{\wedge(N-1)}$$

$(j = 1, \ldots, n)$, $\left(P_j^1 + \tilde{P}_j^1 = I^1\right)$, and

$$\tilde{F}_j^N := \frac{1}{n}\left(I^1 - NP_j^1\right) \wedge I^{\wedge(N-1)} = \frac{1}{n}\tilde{P}_j^{\wedge N}$$

$(j = 1, \ldots, n)$;

$$\sum_{j=1}^{n}\left(F_j^N + \tilde{F}_j^N\right) = \sum_{j=1}^{n}\frac{1}{n}I^{\wedge N} = I^{\wedge N},$$

satisfying all the requirements for normalized positive operator valued measure POV:

**(i)** positivity,

**(ii)** $\sigma$–additivity,

**(iii)** normalisation on the measurable space $(\Omega, \mathcal{A})$, $\Omega = \{0, n\gamma_i\}_{i=1}^n$, the value space of $F^N$.

Then the decomposition (6.1) can be rewritten in the form of a generalized spectral decomposition:

$$NX^1 \wedge I^{\wedge(N-1)} = \sum_{i=1}^{r} n\gamma_i \tilde{F}_i^N + \sum_{k=r+1}^{n} n\gamma_k F_k^N + \sum_{i=r+1}^{n} 0 \cdot \tilde{F}_i^N + \sum_{k=1}^{r} 0 \cdot F_k^N. \tag{6.2}$$

But again, from our operational point of view form (6.1) is more convenient than (6.2), because we are dealing with projection operators $NP_i^1 \wedge \tilde{P}_i^{\wedge(N-1)}$ and $\tilde{P}_i^{\wedge N}$ in (6.1) while $F_i^N$ and $\tilde{F}_i^N$ in (6.2) are not projectors. In this form the positive operator valued measure generated by the family of projectors $\left\{NP_i^1 \wedge \tilde{P}_i^{\wedge(N-1)}, \quad \tilde{P}_i^{\wedge N}\right\}_{i=1}^n$ is not normalized on $\Omega = \{0, \gamma_i\}_{i=1}^n$, $P^N(\Omega) = \sum_{i=1}^n (NP_i^1 \wedge \tilde{P}_i^{\wedge(N-1)} + \tilde{P}_i^{\wedge N}) = nI^{\wedge N}$ $(n = \mathbf{dim}\,\mathcal{H}^1)$,but it provides again a certain probability measure having meaning as it will be seen in Sec. 7.

Now we have three decompositions (actually five of them) of an N-particle antisymmetric 1-body positive-semidefinite operator:

1. the spectral (orthogonal), given by Theorem 3.1,
2. the semi-spectral (non-orthogonal), given by (3.1) (non-normalized), or (3.2) (normalized), and
3. the pseudo-spectral (non-orthogonal) given by Eq. (6.1) (non-normalized), or given by Eq. (6.2) (normalized).

Later on by the pseudo-spectral decomposition we will mean the decomposition (6.1), and we would like to stress once again that the term pseudo-spectral decomposition is introduced to distinguish the decomposition of an N-particle operator $NX^1 \wedge I^{\wedge(N-1)} \geq 0$ that corresponds to the canonical convex decomposition of the 1-particle generating operator $X^1$ into the extreme elements of the cone $\tilde{\mathcal{P}}_N^1$, from all the other generalized spectral decompositions.

In the next section we will be trying to give some physical content to all these three decompositions.

## 7. Physical interpretation

This is only a short attempt of giving a physical interpretation of the obtained herein results, mainly because of the author's very limited knowledge concerning that topic, and because of the need for further, more detailed analysis. The reference books are [2, 10, 23, 34, 35].

For the interpretation we assume that we are able to prepare the system of $N$–fermions in an arbitrary pure state being an element of $\mathcal{H}^{\wedge N}$ (or, in general, in a mixed state represented by a density operator $D^N$), and then physically measure in that state the expectation value of an $N$-particle 1-body operator $NX^1 \wedge I^{\wedge(N-1)}$ representing a certain physical property.

1. Spectral decomposition (Eq.(3.6)):

$$NX^1 \wedge I^{\wedge(N-1)} = \sum_{\mathcal{I}} \lambda_{\mathcal{I}} P^N_{\mathcal{I}}, \tag{7.1}$$

where $P^N_{\mathcal{I}} = N!P^1_{i_1} \wedge \cdots \wedge P^1_{i_j} \wedge P^1_{i_{j+1}} \wedge \cdots \wedge P^1_{i_N}$, $\mathcal{I} = \{i_1, \ldots, i_j, i_{j+1}, \ldots, i_N\}$, $P^N_{\mathcal{I}} P^N_{\mathcal{J}} = \delta_{\mathcal{I}\mathcal{J}} P^N_{\mathcal{I}}$, $\sum_{\mathcal{I}} P^N_{\mathcal{I}} = I^{\wedge N}$, $\lambda_{\mathcal{I}} = \beta_{i_1} + \cdots + \beta_{i_j} + \alpha_{i_{j+1}} + \cdots + \alpha_{i_N}$.

According to the conventional quantum-mechanical interpretation, the set of eigenvalues $\Omega = \{\lambda_{\mathcal{I}}\}_{\mathcal{I}}$ appearing in the spectral decomposition (7.1) has physical meaning, and together with the corresponding projectors $\{P^N_{\mathcal{I}}\}_{\mathcal{I}}$ determines the measure space $(\Omega, \mathcal{A}(\Omega), \mu^P)$, where the measure $\mu^P_{\varphi}(\{\lambda_{\mathcal{I}}\}) =$ **Tr** $(P^N_{\mathcal{I}} P^N_{\varphi})$ gives the probability of getting the eigenvalue $\lambda_{\mathcal{I}}$ in the pure state $\varphi \in \mathcal{H}^{\wedge N}$ characterized by the 1-**dim** projector $P^N_{\varphi}$. In particular if $\varphi$ is an eigenstate $\Phi^N_{\mathcal{J}} = \sqrt{N!}\varphi^1_{j_1} \wedge \cdots \wedge \varphi^1_{j_N}$, $P^N_{\mathcal{J}} = \Phi^N_{\mathcal{J}} \otimes \overline{\Phi}^N_{\mathcal{J}}$, then $\mu^P_{\Phi}$ $(\{\lambda_{\mathcal{I}}\})=$ **Tr** $(P^N_{\mathcal{I}} P^N_{\mathcal{J}})= \delta_{\mathcal{I}\mathcal{J}}$. Therefore, the eigenvalue $\lambda_{\mathcal{I}}$ can be obtained 'experimentally' as the expectation value of the observable $NX^1 \wedge I^{\wedge(N-1)}$ in the pure state $P^N_{\mathcal{I}}$: **Tr** $(NX^1 \wedge I^{\wedge(N-1)} P^N_{\mathcal{I}}) = \lambda_{\mathcal{I}}$. Thus, in the case of the spectral decomposition, the measure space $(\Omega, \mathcal{A}(\Omega), \mu^P)$ has a direct physical meaning.

2. Semi–spectral decomposition (Eq.(3.2)):

$$NX^1 \wedge I^{\wedge(N-1)} = \sum_{i=1}^{s} \beta_i N P^1_i \wedge \tilde{P}^{\wedge(N-1)}_i + \sum_{i=s+1}^{n} \alpha_i N P^1_i \wedge \tilde{P}^{\wedge(N-1)}_i \tag{7.2}$$

Here the values of the set $\Omega = \{\beta_i (i = 1, \ldots, s), \alpha_i (i = s+1, \ldots, n)\}$ cannot be physically measured as the expectation values of the operator $NX^1 \wedge I^{\wedge(N-1)}$ within the state space $\mathcal{H}^{\wedge N}$, and generated by this semi–spectral decomposition measure space $(\Omega, \mathcal{A}(\Omega), \mu)$, has 'no direct' physical meaning as shown in the following. Let again $P^N_{\mathcal{I}}$ be the projection operator onto an eigenfunction $\Phi^N_{\mathcal{I}}$. Then the expectation value

$$\mathbf{Tr}\ \left(NX^1 \wedge I^{\wedge(N-1)} P^N_{\mathcal{I}}\right) = \beta_{i_1} + \cdots + \beta_{i_j} + \alpha_{i_{j+1}} + \cdots + \alpha_{i_N} = \lambda_{\mathcal{I}}$$

and

$$\mu^i_{\mathcal{I}} \equiv \mathbf{Tr}\ \left(N P^1_i \wedge \tilde{P}^{\wedge(N-1)}_i P^N_{\mathcal{I}}\right) = \delta_{i\mathcal{I}} = \begin{cases} 1, & i \in \mathcal{I} \\ 0, & i \notin \mathcal{I} \end{cases}$$

is the probability measure that the number $\beta_i$ (or $\alpha_i$) belonging to the set $\Omega$ will contribute to the physically measured eigenvalue $\lambda_{\mathcal{I}}$. In general, if the system is prepared in the state being a linear combination $\psi^N = \sum_{\mathcal{I}} c_{\mathcal{I}} \Phi^N_{\mathcal{I}}$, $\sum_{\mathcal{I}} \mid c_{\mathcal{I}} \mid^2 = 1$, the probability measure is $\mu^i_\psi = \mathbf{Tr}\ (NP^1_i \wedge \tilde{P}_i^{\wedge(N-1)} P^N_\psi) = \sum_{\mathcal{I}} \mid c_{\mathcal{I}} \mid^2 \delta_{i\mathcal{I}}$, $0 \le \mu^i_\psi \le 1$, $\sum_{i=1}^n \mu^i_\psi = N$, and the expectation value $\mathbf{Tr}\ (NX^1 \wedge I^{\wedge(N-1)} P^N_\psi) = \sum_{i=1}^s \beta_i \mu^i_\psi + \sum_{i=s+1}^n \alpha_i \mu^i_\psi$.

One can get the numbers $\beta_i$ $(i = 1, \ldots, s)$, $\alpha_i$ $(i = s+1, \ldots, n)$ as the expectation values of the operator $NX^1 \wedge I^{\wedge(N-1)}$, extending the underlying Hilbert space (then $I^{\wedge(N-1)}$ refers to the extended space) in order to have spectral decomposition with $\beta_i$ $(i = 1, \ldots, s)$, $\alpha_i$ $(i = s+1, \ldots, n)$ being eigenvalues (Naimark's theorem [1, 10, 23]. But then, the physical situation is changed as the state space is changed, e.g., the positive–semidefinite operator $NX^1 \wedge I^{\wedge(N-1)}$ in the space $\mathcal{H}^{\wedge N}$ will loose this property after the extension of the state space to the properly large one (the $\beta$'s are negative).

3) Pseudo–spectral decomposition Eq.(6.1):

$$NX^1 \wedge I^{\wedge(N-1)} = \sum_{i=1}^{r} \gamma_i \tilde{P}_i^{\wedge N} + \sum_{i=r+1}^{n} \gamma_i N P^1_i \wedge \tilde{P}_i^{\wedge(N-1)}. \tag{7.3}$$

From the physical point of view, a 1–body operator $NX^1 \wedge I^{\wedge(N-1)}$ takes only spin interactions between $N$–electrons into account (Pauli principle). All other 2–body interactions can be treated only approximately by means of the mean field (Hartree–Fock approximation).

Let us give first a physical interpretation to the subspaces of the state space $\mathcal{H}^{\wedge N}$ described by the projection operators $NP^1_i \wedge \tilde{P}_i^{\wedge(N-1)}$ and $\tilde{P}_i^{\wedge N}$ that are expansions to the N–particle space of the extreme elements of the polar cone $\tilde{\mathcal{P}}^1_N$.

The operator $NP^1_i \wedge \tilde{P}_i^{\wedge(N-1)}$ projects onto the subspace of $\mathcal{H}^{\wedge N}$ consisting of the antisymmetric functions of the type $\varphi^1_i \wedge \psi^{N-1}_{\tilde{i}}$, where $\psi^{N-1}_{\tilde{i}}$ is strongly orthogonal to $\varphi^1_i$ $(\psi^{N-1}_{\tilde{i}} \underline{\perp} \varphi^1_i)$, i.e., $\varphi^1_i$ is not occupied (does not appear) when $\psi^{N-1}_{\tilde{i}}$ is expanded in the orthonormal basis consisting of the $(N-1)$–particle determinantal functions build up from the 1–particle complete orthonormal system $\{\varphi^1_i\}_{i=1}^n$. From the physical point of view, any function belonging to this subspace describes a state of $N$ electrons (fermions) in which one electron must be in a 1-particle state $\varphi^1_i$ while the remaining $N-1$ electrons are in arbitrary state (in general it could be a correlated state, i.e., not of the determinantal form $\sqrt{(N-1)!}\varphi^1_{i_2} \wedge \ldots \wedge \varphi^1_{i_N}$). There is no correlation (except spin) between one electron and the remaining $N-1$ electrons. So, in this subspace we can (and must) speak about the individual state of one electron at least ('a particle state' in the second quantization language).

On the other hand the operator $\tilde{P}_i^{\wedge N}$ projects onto the subspace of $\mathcal{H}^{\wedge N}$ consisting of antisymmetric functions $\psi^N_{\tilde{i}}$ not containing $\varphi^1_i$ in their expansion in the complete orthonormal determinantal basis $\sqrt{N!}\varphi^1_{i_1} \wedge \ldots \wedge \varphi^1_{i_N}$ $(i_1 < \ldots < i_N)$, $\varphi^1_{i_k} \in \{\varphi^1_i\}_{i=1}^n$, i.e., $\psi^N_{\tilde{i}} \underline{\perp} \varphi^1_i$. Physically, these functions describe in general a correlated state of $N$–electrons in which must be a hole, i.e., no one electron may occupy the state $\varphi^1_i$ (a 'hole state' in the second quantization language). Obviously a determinantal state not containing $\varphi^1_i$ also belongs to this subspace, but there are many correlated states there as well provided that the dimension of the subspace is appropriate.

Thus, from the physical point of view the pseudo–spectral decomposition (7.3) introduces in the state space $\mathcal{H}^{\wedge N}$ a classification of states into two types:

**(i)** belonging to the subspace described by the projection operator $NP_i^1 \wedge \tilde{P}_i^{\wedge(N-1)}$ $(i = 1, \ldots, n)$ (a 'particle state'); in this subspace the existence of a completely correlated state of the system of $N$–fermions is impossible (a 'normal state'),

**(ii)** belonging to the subspace $\tilde{P}_i^{\wedge N}$ $(i = 1, \ldots, n)$ (a 'hole' state); in this subspace a completely correlated state of $N$–fermions is available, and then in this state the $N$ electrons must be treated as a bulk (a 'collective state').

We observe that the AGP–function [5] $\left(g^2\right)^{\wedge \frac{N}{2}}$ describing a superconducting state of $N$ fermions ($N$ even) in the BCS model is of the type **(ii)**.

To have physical interpretation of the measure space $(\Omega, \mathcal{A}, \mu)$ generated by the pseudo–spectral decomposition we take, as in the case 2), the expectation value of (7.3) in the eigenstate $P_{\mathcal{I}}^N$:

$$\begin{aligned}
\mathbf{Tr}\,\left(NX^1 \wedge I^{\wedge(N-1)} P_{\mathcal{I}}^N\right) &= \sum_{i=1}^{r} \gamma_i \,\mathbf{Tr}\,\left(\tilde{P}_i^{\wedge N} P_{\mathcal{I}}^N\right) + \sum_{i=r+1}^{n} \gamma_i \,\mathbf{Tr}\,\left(NP_i^1 \wedge \tilde{P}_i^{\wedge(N-1)} P_{\mathcal{I}}^N\right) \\
&= \sum_{i=1}^{r} \gamma_i \left(1 - \delta_{i\mathcal{I}}\right) + \sum_{i=r+1}^{n} \gamma_i \delta_{i\mathcal{I}} \\
&= \sum_{i=1}^{r} \gamma_i \tilde{\mu}_{\mathcal{I}}^i + \sum_{i=r+1}^{n} \gamma_i \mu_{\mathcal{I}}^i = \lambda_{\mathcal{I}}
\end{aligned}$$

(to get the last equality requires substitution for $\gamma$'s their definitions, Eqs.(5.5)–(5.7),and then some calculations). Here $0 \leq \mu_i \leq 1$, $0 \leq \tilde{\mu}_i \leq 1$, $(i = 1, \ldots, n)$ and $\sum_{i=1}^n (\tilde{\mu}_i + \mu_i) = n = \mathbf{dim}\,\mathcal{H}^1$. Thus we see that the set $\Omega$ consists of the numbers $\{\gamma_i\}_{i=1}^n$, and again $\mu_{\mathcal{I}}^i$ is the probability that the number $\gamma_i$ $(i \in \{r+1, \ldots, n\})$ contributes to the eigenvalue $\lambda_{\mathcal{I}}$, and this contribution like in Eq.(7.2) comes from the 'particle', while, $\tilde{\mu}_{\mathcal{I}}^i$ is the probability of the contribution of number $\gamma_i$ $(i \in \{1, \ldots, r\})$ to the eigenvalue $\lambda_{\mathcal{I}}$, but this contribution comes from the 'hole' state. Similarly to the semi–spectral decomposition (7.2), the numbers $\gamma_i \in \Omega$ are not in general eigenvalues of $NX^1 \wedge I^{\wedge(N-1)} \geq 0$ and therefore are not physically measurable themselves. Nonetheless, under some conditions this may happen.

**Theorem 7.1:** *Let a* 1*-body positive–semidefinite operator* $NX^1 \wedge I^{\wedge(N-1)}$ *has the following pseudo-–spectral decomposition*

$$NX^1 \wedge I^{\wedge(N-1)} = \sum_{i=1}^{r} \gamma_i \tilde{P}_i^{\wedge N} + \sum_{i=r+1}^{m} \gamma_i NP_i^1 \wedge \tilde{P}_i^{\wedge(N-1)} + \sum_{k=m+1}^{n} 0 \cdot NP_k^1 \wedge \tilde{P}_k^{\wedge(N-1)}, \tag{7.4}$$

*where* $n - m \geq N - r + 1$*, then*

**(i)** *zero is the lowest eigenvalue of the operator* $NX^1 \wedge I^{\wedge(N-1)}$ *and*

$$\textbf{\textit{dim ker}}\, NX^1 \wedge I^{\wedge(N-1)} = \begin{pmatrix} n-m \\ N-r \end{pmatrix},$$

**(ii)** *the* $\gamma_i$ $(i = 1, \ldots, m)$ *are the* $m$ *lowest positive eigenvalues of the operator* $NX^1 \wedge I^{\wedge(N-1)}$*, and all the other (higher) eigenvalues are expressible as sums of the lowest ones,*

**(iii)** *the 1–particle operator* $X^1$ *that leads to (7.4) has the following spectral decomposition*

$$X^1 = \sum_{i=1}^{s} \beta_i P_i^1 + \sum_{i=s+1}^{r} \alpha_i P_i^1 + \sum_{k=m+1}^{n} -t_N(r) P_k^1, \tag{7.5}$$

*where*

$$t_N(r) = (N-r)^{-1} \left( \sum_{i=1}^{s} \beta_i + \sum_{i=s+1}^{r} \alpha_i \right),$$

*while*

$$\gamma_i = -\left(t_N(r) + \beta_i\right), \quad (i = 1, \ldots, s), \tag{7.6a}$$

$$\gamma_i = -\left(t_N(r) + \alpha_i\right) \quad (i = s+1, \ldots, r), \tag{7.6b}$$

*and*

$$\gamma_i = t_N(r) + \alpha_i \quad (i = r+1, \ldots, n). \tag{7.6c}$$

Proof. Let us recall that the projection operator onto the nullspace of the projector $NP_i^1 \wedge \tilde{P}_i^{\wedge(N-1)}$ is equal to $\tilde{P}_i^{\wedge N}$ and vice versa, i.e., $NP_i^1 \wedge \tilde{P}_i^{\wedge(N-1)} + \tilde{P}_i^{\wedge N} = I^{\wedge N}$ $(i = 1, \ldots, n)$, and that the eigenfunctions of $NX^1 \wedge I^{\wedge(N-1)}$ are determinantal states $\Phi_{\mathcal{I}}^N = \sqrt{N!}\left(\varphi_{i_1}^1 \wedge \cdots \wedge \varphi_{i_N}^1\right)$, $\varphi_i^1 \in \mathcal{H}^1$ $(i = 1, \ldots, n)$, $\mathcal{I} = \{i_1, \ldots, i_N\}$ $(i_k < i_{k+1})$, with $\varphi_i^1 \otimes \overline{\varphi}_i^1 = P_i^1$. We denote by $P_{\mathcal{I}}^N$ the projection operator onto the eigenfunctions $\Phi_{\mathcal{I}}^N$. The proof consists in calculating the expectation values of $NX^1 \wedge I^{\wedge(N-1)}$ in the appropriate eigenstates. To show that zero is the lowest eigenvalue of (7.4) we choose an eigenstate of the form $\Phi_{\mathcal{K}}^N = \sqrt{N!}\left(\varphi_i^1 \wedge \cdots \wedge \varphi_r^1 \wedge \varphi_{k_{r+1}}^1 \wedge \cdots \wedge \varphi_{k_N}^1\right)$, where $\left\{k_{r+1}, \ldots, \varphi_{k_N}^1\right\} \subset \{m+1, \ldots, n\}$, then due to the remainder $\mathbf{Tr}\left(NX^1 \wedge I^{\wedge(N-1)} P_{\mathcal{K}}^N\right) = 0$, and because the 1–particle functions $\varphi_1^1, \ldots, \varphi_r^1$ must always be kept in the determinant belonging to the kernel, while the $\varphi_{r+1}^1, \ldots, \varphi_m^1$ cannot appear, the dimension of the kernel is equal $\begin{pmatrix} n-m \\ N-r \end{pmatrix}$. Now we show that $\gamma_i$ $(i = 1, \ldots, r)$ are eigenvalues of $NX^1 \wedge I^{\wedge(N-1)}$ belonging to the eigenstates $\Phi_{\tilde{i}}^N = \sqrt{N!}\left(\varphi_1^1 \wedge \cdots \wedge \varphi_{i-1}^1 \wedge \varphi_{i+1}^1 \wedge \varphi_r^1 \wedge \varphi_{k_{r+1}}^1 \wedge \cdots \wedge \varphi_{k_N}^1\right)$. We have $\mathbf{Tr}\left(\tilde{P}_j^{\wedge N} P_{\tilde{i}}^N\right) = \delta_{ij}$ $(j = 1, \ldots, r)$, and $\mathbf{Tr}\left(NP_j^1 \wedge \tilde{P}_j^{\wedge(N-1)} P_{\tilde{i}}^N\right) = 0$, $(j = r+1, \ldots, m)$. Hence the expectation value $\mathbf{Tr}\left(NX^1 \wedge I^{\wedge(N-1)} P_{\tilde{i}}^N\right) = \gamma_i$ $(i = 1, \ldots, r)$. Therefore $\gamma_i$ $(i = 1, \ldots, r)$ are eigenvalues of $NX^1 \wedge I^{\wedge(N-1)}$. Similarly, taking $\Phi_i^N = \sqrt{N!}\left(\varphi_1^1 \wedge \cdots \wedge \varphi_r^1 \wedge \varphi_i^1 \wedge \varphi_{k_{r+2}}^1 \wedge \cdots \wedge \varphi_{k_N}^1\right)$, $i \in \{r+1, \ldots, m\}$, $\{k_{r+2},\ldots,k_N\} \subset \{m+1,\ldots,n\}$, we have $\mathbf{Tr}\left(NX^1 \wedge I^{\wedge(N-1)}P_i^N\right) = \gamma_i$ $(i = r+1,\ldots,m)$, and therefore also $\gamma_i$ $(i = r+1,\ldots,m)$ are eigenvalues of $NX^1 \wedge I^{\wedge(N-1)}$. Finally taking the expectation value of $NX^1 \wedge I^{\wedge(N-1)}$ in all the other eigenstates, being $N$–particle determinantal functions formed from the 1–particle basis $\left\{\varphi_i^1\right\}_{i=1}^n$, we obtain all the other eigenvalues as sums of the eigenvalues 0, $\gamma_i$ $(i = 1,\ldots,m)$, as follows from the r.h.s of (7.4). Straightforward from Theorems 6.1 and 5.1 we can see that (7.5) implies (7.4) with $\gamma$'s given by Eq (7.6). This completes the proof. □

As it follows from the Theorem and its proof it is worthwhile to observe that the decomposition (7.4) through (7.5) divides the 1–particle Hilbert space $\mathcal{H}^1$ into three mutually orthogonal subspaces $\mathcal{H}^1 = \mathcal{H}_i^1 \oplus \mathcal{H}_j^1 \oplus \mathcal{H}_k^1$ being spanned by the following orthonormal basis appropriately : $\left\{\varphi_i^1\right\}_{i=1}^r$, $\left\{\varphi_j^1\right\}_{j=r+1}^m$, $\left\{\varphi_k^1\right\}_{k=m+1}^n$. Constructing determinantal eigenfunctions of the operator $NX^1 \wedge I^{\wedge(N-1)}$ we use elements of the basis from different subspaces depending on whether the eigenfunction belongs to the kernel: $\mid 1,\ldots,r;k_{r+1}\ldots k_N\rangle$, 'hole state' : $\mid 1,\ldots,\tilde{i},\ldots,r;k_{r+1},\ldots,k_i,\ldots,k_N\rangle$ ($\tilde{i}$ means 'no i'), or 'particle state': $\mid 1,\ldots,r;j;k_{r+2},\ldots k_N\rangle$.

Assuming the above physical interpretation of the pseudo–spectral decomposition we may try to analyze the behaviour of $N$ fermions described by the '1–body Hamiltonian' $NX^1 \wedge I^{\wedge(N-1)}$ given by (7.4). In this paper the fermion 1–body operator $NX^1 \wedge I^{\wedge(N-1)}$ is always positive–semidefinite because we are interested in the dual cone $\tilde{\mathcal{P}}_N^1$ of the set of fermion $N$–representable 1–particle density operators $\tilde{P}_N^1$. Therefore zero is the lowest available eigenvalue and then the nullspace $\mathbf{ker}\, NX^1 \wedge I^{\wedge(N-1)}$ is the ground state subspace. Also the pseudo–spectral decomposition (7.3) (and (7.4)) refers to this situation, hence, $\gamma$'s are non–negative. In a more realistic physical situation we would rather have to take into account also the negative cone. Treating (7.4) as a 'model Hamiltonian' we might think of the eigenvalue zero of (7.4) as of the 'relative zero', i.e., the lowest eigenvalue of the system of $N$–fermions, and the nullspace as the lowest eigenspace. Also $N$ would rather refer only to the part of the electrons in our system, that could be treated in the 1–body approximation represented by (7.4) (e.g., the electrons in the conducting zone, say). In this case the identity operator in (7.4) would be the projection operator onto the subspace under consideration. What we would like to demonstrate is how the splitting of the state space into two physically rather different classes given by the pseudo–spectral decomposition of a fermion $N$–particle 1–body operator could help in setting up a mathematical model describing the appearance in the system of $N$–fermions a 'collective' or 'normal' first excited state.

The assumptions of a mathematical model describing both normal and collective behaviour of a system of $N$–fermions:

**$\mathbf{a}_1$)** A model $N$–particle 1–body fermion Hamiltonian $NX^1 \wedge I^{\wedge(N-1)}$ possesses both collective and normal eigenvalues and corresponding collective and normal eigenfunctions.

**$\mathbf{a}_2$)** The collective eigenstates must belong to the subspaces $\tilde{P}_i^{\wedge N}$ $(i = 1,\ldots,r)$, Eq. (7.4), i.e., they are 'hole' states, with the corresponding 'collective eigenvalues' $\gamma_i$ $(i = 1,\ldots,r)$, while the states belonging to the subspaces $NP_j^1 \wedge \tilde{P}_j^{\wedge(N-1)}$ $(j = r+1,\ldots,m)$ are 'normal' states, i.e, 'particle states', with the corresponding 'normal eigenvalues' $\gamma_j$ $(j = r+1,\ldots,m)$.

**$\mathbf{a}_3$)** Zero is the bottom eigenvalue, and $\mathbf{ker}\, NX^1 \wedge I^{\wedge(N-1)}$ is the bottom eigenspace.

We formulate the necessary conditions for the lowest excited state being a collective one:

$s_1$) The collective eigenstate must be as much correlated as possible.

$s_2$) The normal state must be as much uncorrelated as possible.

$s_3$) The lowest eigenvalue corresponding to the collective state should be below the lowest eigenvalue corresponding to the normal eigenstate. In other words it must be a gap between the bottom eigenvalue and the first excited eigenvalue corresponding to the normal state within which lies the eigenvalue corresponding to the collective state (the larger the difference the higher the temperature in which the collective state is stable).

Now we want to construct a 1–body Hamiltonian $NX^1 \wedge I^{\wedge(N-1)}$ satisfying the above necessary conditions, and then find the corresponding 1–particle operator $X^1$, i.e., its spectral decomposition.

Condition $s_1$) requires a high degeneracy of the appropriate collective eigenvalue $\gamma_i$ $(i = 1, \ldots, r)$ which can be achieved on one hand by the condition $\kappa \equiv n - m > N - r + 1$ (or even $\gg$), on the other by setting some of the $\gamma_i$ $(i = 1, \ldots, r)$ equal one to each other, which might considerably increase the dimension of the corresponding eigenspace as $r$ is of the order of $N$ $(r \leq N - 1)$ but which constrains the number of different 'collective eigenvalues'.

Condition $s_2$) requires all $\gamma_j$ $(j = r+1, \ldots, m)$ to be different one from each other, then reasonably small $r$ and $\kappa \equiv n - m$ (but $r + \kappa > N + 1$).

Condition $s_3$) needs $\gamma_r < \gamma_{r+1}$ (as we have in general $\gamma_1 \geq \gamma_2 \geq \ldots \geq \gamma_r$, and $\gamma_{r+1} \leq \gamma_{r+2} \leq \ldots \leq \gamma_m$, all $\gamma$'s positive ). We consider two simple cases of our mathematical model for collective behaviour of a system of $N$ fermions :

**(i)** one 'collective eigenvalue', i.e., all $\gamma_i$ $(i = 1, \ldots .r)$ equal one to each other; this would correspond to the collective behaviour of type I,

**(ii)** two collective eigenvalues within the gap, i.e., $\gamma_1 = \cdots = \gamma_s$, and $\gamma_{s+1} = \cdots = \gamma_r$; this case would correspond to collective behaviour of type II.

(In a more physically realistic model the sharp levels probably should be rather diffused to bands, which could be in principle done by arranging many eigenvalues lying close together.)

**(i)** Type I collective behaviour: We take the 1–particle operator $X^1$, Eq. (7.5), of the form:

$$X^1 = \sum_{i=1}^{\frac{N}{2}} \beta P_i^1 + \sum_{j=\frac{N}{2}+1}^{m} \alpha_j P_j^1 + \sum_{k=m+1}^{n} -\beta P_k^1, \tag{7.7}$$

with $\kappa \equiv n - m \geq \frac{N}{2} + 1$ (we assume $N$ is even; for $N$ odd the sum will be until $r = \frac{N+1}{2}$, while

$\kappa \geq \frac{N+1}{2}$). This leads to the following $N$–particle 1–body 'Hamiltonian', Eq. (7.4),:

$$NX^1 \wedge I^{\wedge(N-1)} = \sum_{i=1}^{\frac{N}{2}} -2\beta \tilde{P}_i^{\wedge N} + \sum_{j=\frac{N}{2}+1}^{m} (\beta + \alpha_j)\, NP_j^1 \wedge \tilde{P}_j^{\wedge(N-1)} + \sum_{k=m+1}^{n} 0 \cdot NP_k^1 \wedge \tilde{P}_k^{\wedge(N-1)}, \quad (7.8)$$

where $\beta + \alpha_j > -2\beta$, $\beta < 0$, $\alpha_j > 0$ ($j = \frac{N}{2}+1, \ldots, m$), i.e., condition $s_3$) must be satisfied. Condition $s_2$) is satisfied provided all $\alpha_j$ ($j = \frac{N}{2}+1, \ldots, m$) are different, and then the corresponding eigenspaces have dimension $\begin{pmatrix} \kappa \\ \frac{N}{2}-1 \end{pmatrix}$. Since we have all $\gamma_i = -2\beta$ $\left(i = 1, \ldots, \frac{N}{2}\right)$ equal one to each other, the corresponding 'collective eigenvalue' $(-2\beta)$ has degeneracy $\frac{N}{2} \begin{pmatrix} \kappa \\ \frac{N}{2} \end{pmatrix}$, and in this eigenspace there exist $N$–electron completely correlated (i.e., no 1–particle Grassmann factors) eigenfunctions, which guarantees $s_1$) being satisfied. Thus, according to our model, operator (7.8) possesses the appropriate structure of eigenvalues and eigenfunctions in order to describe collective phenomena. The diagram visualizing the situation is placed in Appendix 3. The arrows indicate the contribution of the 1–particle eigenvalues to the $N$–particle ones. The bottom eigenvalue of the $N$–particle operator $NX^1 \wedge I^{\wedge(N-1)}$, Eq. (7.8), is zero $\left(\frac{N}{2}\beta + \frac{N}{2}(-\beta)\right)$, and the corresponding ground eigenspace of the dimension $\begin{pmatrix} \kappa \\ \frac{N}{2} \end{pmatrix}$ is span by the eigenfunctions of the type

$$\psi^N_{ground} = \sum_{\mathcal{K}} c_{\mathcal{K}} \sqrt{N!} \varphi_1^1 \wedge \ldots \wedge \varphi^1_{\frac{N}{2}} \wedge \varphi^1_{k_{\frac{N}{2}+1}} \wedge \ldots \wedge \varphi^1_{k_N} \equiv \sum_{\mathcal{K}} c_{\mathcal{K}} \mid 1, \ldots, \tfrac{N}{2}; k_{\frac{N}{2}+1}, \ldots, k_N \rangle,$$

where $\mathcal{K}$ describes the $\begin{pmatrix} \kappa \\ \frac{N}{2} \end{pmatrix}$ configurations of the $N$–particle determinants with $\frac{N}{2}$ fixed 1–particle functions, and the remaining $\frac{N}{2}$ functions $\varphi^1_k$ chosen from the set $\left\{\varphi^1_k\right\}^n_{k=m+1}$. Thus, an element of the ground subspace is largely uncorrelated as having in general $\frac{N}{2}$ 1–particle Grassmann factors, i.e., $\psi^N = \varphi^1_1 \wedge \ldots \wedge \varphi^1_{\frac{N}{2}} \wedge \psi^{\frac{N}{2}}$. The first excited normal eigenstate belonging to the eigenvalue $\beta + \alpha_{\frac{N}{2}+1}$ differs from the ground state by replacing one electron corresponding to the eigenvalue $\alpha_{\frac{N}{2}+1}$ (the lowest unoccupied). The degeneracy is $\begin{pmatrix} \kappa \\ \frac{N}{2}-1 \end{pmatrix}$, and the corresponding eigenfunctions are of the form

$$\psi^N_{normal} = \sum_{\mathcal{K}} c_{\mathcal{K}} \sqrt{N!} \varphi_1^1 \wedge \ldots \wedge \varphi^1_{\frac{N}{2}+1} \wedge \varphi^1_{k_{\frac{N}{2}+2}} \wedge \ldots \wedge \varphi^1_{k_N} \equiv \sum_{\mathcal{K}} \mid 1, \ldots, \tfrac{N}{2}+1; k_{\frac{N}{2}+2}, \ldots, k_N \rangle.$$

These eigenfunctions are also very uncorrelated because they have $\frac{N}{2}+1$ 1–particle Grassmann factors: $\psi^N_{normal} = \varphi^1_1 \wedge \ldots \wedge \varphi^1_{\frac{N}{2}+1} \wedge \psi^{\frac{N}{2}-1}$. Above the first excited $N$–particle normal state there is an array of higher normal states corresponding to higher 1–particle excitations (i.e., instead of $\alpha_{\frac{N}{2}+1}$ we substitute in general $\alpha_j$, $j = \frac{N}{2}+2, \ldots, m$). These states are not indicated in the picture. Within the gap between zero and $\beta + \alpha_{\frac{N}{2}+1}$ we have a'collective eigenvalue' $(= -2\beta)$ with corresponding 'collective eigenspace' of dimension $\frac{N}{2}\left(\begin{array}{c}\kappa\\ \frac{N}{2}+1\end{array}\right)$. In this subspace there are completely correlated (no 1–particle Grassmann factor) eigenfunctions describing collective behaviour. They are of the following form:

$$\begin{aligned}\psi^N_{collective} &= \sum_{i,\mathcal{K}} c_{i\mathcal{K}}\sqrt{N!}\varphi^1_1\wedge\ldots\wedge\varphi^1_{i-1}\wedge\varphi^1_{i+1}\wedge\ldots\wedge\varphi^1_{\frac{N}{2}}\wedge\varphi_{k_{\frac{N}{2}+1}}\wedge\ldots\wedge\varphi^1_{k_i}\wedge\ldots\wedge\varphi^1_{k_N}\\ &\equiv \sum_{i,\mathcal{K}} c_{i\mathcal{K}}\mid 1,\ldots,\tilde{i},\ldots,\tfrac{N}{2};k_{\frac{N}{2}+1},\ldots,k_i,\ldots,k_N,\rangle\end{aligned}$$

where $i$ runs from 1 to $\frac{N}{2}$ ($\tilde{i}$ means 'no i'), while $\mathcal{K}$ describes the $\left(\begin{array}{c}\kappa\\ \frac{N}{2}+1\end{array}\right)$ configurations of $N$–particle determinants with $\frac{N}{2}-1$ fixed 1–particle functions (for each $i=1,\ldots,\frac{N}{2}$), and the remaining $\frac{N}{2}+1$ functions choosen from the set: $\{\varphi^1_k\}^n_{k=m+1}$.

The 1–particle functions $\varphi^1_i$ ($i=1,\ldots,n$) which are eigenfunctions of the 1–particle operator $X^1$, Eq. (7.7), contain both spatial and spin variables (spin–orbitals), and we may assign 'spin down' for $i=1,\ldots,\frac{N}{2}$, and 'spin up' for $i=\frac{N}{2}+1,\ldots,n$ (we assume $N$ is even). Then we have some kind of pairing like in the BCS model for superconductivity. Under this assumption the bottom and normal states will be singlets, while the collective state a triplet state, and therefore metastable. For $N$ odd all the states are doublets, and therefore the collective eigenstate is not metastable. Changing the spin orientation one may get the collective state with a large magnetic moment which might have some relation to the ferromagnetic state.

Now we arrange two 'collective eigenvalues' within the gap by setting $\gamma_1=\cdots=\gamma_s$, and $\gamma_{s+1}=\cdots=\gamma_r$. Perhaps, we could have a model for superconductor of type II (type I would correspond to one collective eigenvalue if we set $\mid\beta\mid=\Delta=\frac{3}{2}kT_c$), or for a two phase magnetic behaviour.

**(ii)** Type II collective behaviour: We take

$$X^1=\sum_{i=1}^{\frac{N}{4}}2\beta P^1_i+\sum_{i=\frac{N}{4}+1}^{\frac{N}{2}}2\alpha P^1_i+\sum_{j=\frac{N}{2}+1}^{m}\alpha_j P^1_j+\sum_{k=m+1}^{n}-(\beta+\alpha)P^1_k,$$

which leads to

$$\begin{aligned}NX^1\wedge I^{\wedge(N-1)} &= \sum_{i=1}^{\frac{N}{4}}-(3\beta+\alpha)\tilde{P}^{\wedge N}_i+\sum_{i=\frac{N}{4}+1}^{\frac{N}{2}}-(\beta+2\alpha)\tilde{P}^{\wedge N}_i\\ &+ \sum_{j=\frac{N}{2}+1}^{m}(\beta+\alpha+\alpha_j)NP^1_j\wedge\tilde{P}^{\wedge(N-1)}_j+\sum_{k=m+1}^{n}0\cdot NP^1_k\wedge\tilde{P}^{\wedge(N-1)}_k.\end{aligned}$$

The requirement for having both collective eigenvalues within the gap is $4\beta+2\alpha+\alpha_j>0$ ($j=\frac{N}{2}+1,\ldots,m$). The diagram representing mutual interrelation between the eigenvalues is placed also in Appendix 3. The corresponding eigenfunctions for the ground and normal excited state are of the same form as before, while

$$\psi^N_{collective\,1} = \sum_{i=1}^{\frac{N}{4}} \sum_{\mathcal{K}} c_{i\mathcal{K}} \mid 1, \ldots, \tilde{i}, \ldots, \tfrac{N}{4}, \ldots, \tfrac{N}{2}; k_{\frac{N}{2}+1}, \ldots, k_i, \ldots, k_N\rangle,$$

$$\psi^N_{collective\,2} = \sum_{i=\frac{N}{4}+1}^{\frac{N}{2}} \sum_{\mathcal{K}} c_{i\mathcal{K}} \mid 1, \ldots, \tfrac{N}{4}, \ldots, \tilde{i}, \ldots, \tfrac{N}{2}; k_{\frac{N}{2}+1}, \ldots, k_i, \ldots, k_N\rangle.$$

Thus, the mathematical model we have been considering seems to be rather flexible and, depending on the expert's opinion, perhaps could be adjusted to some real physical situations in which collective phenomena are involved (superconductivity, magnetic phenomena, collective states of nuclei). However, the limitation is that we have only a fermion '1–body Hamiltonian', and the appearance of collective behaviour is due to the Pauli principle. All other physical interactions like Coulomb repulsion between electrons could be taken into account only through the 1–particle operator $X^1$ by means of the mean field approximation. Solution of the fermion $N$–representability problem for a 2–particle density operator perhaps would have helped if it had been known.

**Acknowledgements**

The author would like to thank Professor Brian G. Wybourne for the fruitful discussions in the course of preparing the manuscript, its careful reading, and the improvements. I am also thankfull to my son Karol for devoting much of his time for typing the manuscript in LaTeX.

## Appendix 1

**Proof of Theorem 3.1.** The proof consists in straightforward decomposition of $NX^1 \wedge I^{\wedge\,(N-1)}$ using Lemma 3.1. In the following $P^1_{1:s} \equiv \sum_{i=1}^{s} P^1_i$, $\tilde{P}^1_{1:s} \equiv I^1 - P^1_{1:s}$, where $I^1 = \sum_{i=1}^{n} P^1_i$ is the resolution of the identity operator $I^1$ on the 1–particle Hilbert space $\mathcal{H}^1$ onto the mutually orthogonal 1–dimensional projection operators $P^1_i$ $(i = 1 \ldots n)$. While,

$$I^{\wedge(N-1)} = (P^1_{1:s} + \tilde{P}^1_{1:s})^{\wedge(N-1)} = \sum_{j=0}^{N-1} \binom{N-1}{j} P^{\wedge j}_{1:s} \wedge \tilde{P}^{\wedge(N-1-j)}_{1:s}.$$

We have,

$$\begin{aligned} NX^1 \wedge I^{\wedge(N-1)} &= \left(\sum_{i=1}^{s} \beta_i P^1_i + \sum_{i=s+1}^{n} \alpha_i P^1_i\right) \wedge \sum_{j=0}^{N-1} N \binom{N-1}{j} P^{\wedge j}_{1:s} \wedge \tilde{P}^{\wedge(N-1-j)}_{1:s} \\ &= \sum_{j=0}^{N-1} \binom{N}{j} (N-j) \sum_{i=1}^{s} \beta_i P^1_i \wedge P^{\wedge j}_{1:s} \wedge \tilde{P}^{\wedge(N-1-j)}_{1:s} \end{aligned}$$

$$
\begin{aligned}
&+ \sum_{j=0}^{N-1} \binom{N}{j} (N-j) P_{1:s}^{\wedge j} \wedge \sum_{i=s+1}^{n} \alpha_i P_i^1 \wedge \tilde{P}_{1:s}^{\wedge (N-1-j)} \\
&= \sum_{j=0}^{N} \binom{N}{j} \sum_{i=1}^{s} \beta_i j P_i^1 \wedge P_{1:s}^{\wedge (j-1)} \wedge \tilde{P}_{1:s}^{\wedge (N-j)} \\
&+ \sum_{j=0}^{N} \binom{N}{j} P_{1:s}^{\wedge j} \wedge \sum_{i=s+1}^{n} \alpha_i (N-j) P_i^1 \wedge \tilde{P}_{1:s}^{\wedge (N-j-1)}. \qquad \text{(A1.1)}
\end{aligned}
$$

Now, we observe that

$$
\begin{aligned}
\sum_{i=1}^{s} \beta_i j P_i^1 \wedge P_{1:s}^{\wedge (j-1)} &= \sum_{i_1=1}^{s} \ldots \sum_{i_j=1}^{s} (\beta_{i_1} + \cdots + \beta_{i_j}) P_{i_1}^1 \wedge \ldots \wedge P_{i_j}^1 \qquad \text{(A1.2)} \\
&= \sum_{1 \le i_1 < \ldots < i_j \le s} (\beta_{i_1} + \cdots + \beta_{i_j}) j! P_{i_1}^1 \wedge \ldots \wedge P_{i_j}^1 ,
\end{aligned}
$$

where $j!P_{i_1}^1 \wedge \ldots \wedge P_{i_j}^1$ is a 1–**dim** projector on $\mathcal{H}^{\wedge j}$. Similarly,

$$
\begin{aligned}
\sum_{i=s+1}^{n} \alpha_i (N-j) P_i^1 \wedge \tilde{P}_{1:s}^{\wedge (N-j-1)} &= \sum_{s+1 \le i_{j+1} < \ldots < i_N \le n} (\alpha_{i_{j+1}} + \cdots + \alpha_{i_N}) \qquad \text{(A1.3)} \\
&\times \; (N-j)! P_{i_{j+1}}^1 \wedge \ldots \wedge P_{i_N}^1 .
\end{aligned}
$$

Taking into account that

$$
P_{1:s}^{\wedge j} = \sum_{i_1=1}^{s} P_{i_1}^1 \wedge \ldots \wedge \sum_{i_j=1}^{s} P_{i_j}^1 = \sum_{1 \le i_1 < \ldots < i_j \le s} j! P_{i_1}^1 \wedge \ldots \wedge P_{i_j}^1 \ ,
$$

and

$$
\tilde{P}_{1:s}^{\wedge (N-j)} = \sum_{i=s+1}^{n} P_{i_{j+1}}^1 \wedge \ldots \wedge \sum_{i_N=s+1}^{n} P_{i_N}^1 = \sum_{s+1 \le i_{j+1} < \ldots < i_N \le n} (N-j)! P_{i_{j+1}}^1 \wedge \ldots \wedge P_{i_N}^1
$$

we arrive from (A1.1)–(A1.3) at the following spectral decomposition of the 1–body operator

$$
\begin{aligned}
N X^1 \wedge I^{\wedge (N-1)} &= \sum_{j=0}^{N} \Bigg( \sum_{1 \le i_1 < \ldots < i_j \le s} \; \sum_{s+1 \le i_{j+1} < \ldots < i_N \le n} (\beta_{i_1} + \cdots + \beta_{i_j} + \\
&+ \; \alpha_{i_{j+1}} + \cdots + \alpha_{i_N}) N! P_{i_1}^1 \wedge \ldots \wedge P_{i_j}^1 \wedge P_{i_{j+1}}^1 \wedge \ldots \wedge P_{i_N}^1 \Bigg), \qquad \text{(A1.4)}
\end{aligned}
$$

since $\sum_{1 \le i_1 < \ldots < i_N \le n} N! P_{i_1}^1 \wedge \ldots \wedge P_{i_N}^1 = I^{\wedge N}$ is the resolution of the identity $I^{\wedge N}$ on $\mathcal{H}^{\wedge N}$ onto the mutually orthogonal 1–**dim** projectors.This proves the first part of the theorem.

It is known that a self–adjoint operator is positive–semidefinite if and only if its eigenvalues are non–negative. Hence all the eigenvalues

$$\beta_{i_1} + \cdots + \beta_{i_j} + \alpha_{i_{j+1}} + \cdots + \alpha_{i_N}(1 \leq i_1 < \ldots < i_j \leq s),\ s+1 \leq i_{j+1} < \ldots < i_N \leq n)$$

in (A1.4) must be non–negative. This completes the proof. □

**Appendix 2**

**Proof of Lemma 4.1** (this is a '1–body' version of the proof for $\tilde{\mathcal{P}}_N^2$ given in [21, p. 20]).

Suppose $X^1 \in \tilde{\mathcal{P}}_N^1$ is extreme and **ker** $\Gamma_1^N X^1$ is not maximal, i.e., there exists $\Gamma_1^N X_1^1 \geq 0$ such that **ker** $\Gamma_1^N X^1 \subset$ **ker** $\Gamma_1^N X_1^1$. Then, $\epsilon > 0$ can be chosen in such a way that $\Gamma_1^N X^1 - \epsilon\, \Gamma_1^N X_1^1 = \Gamma_1^N X_2^1 \geq 0$. Thus, $\Gamma_1^N X^1 = \epsilon\, \Gamma_1^N X_1^1 + \Gamma_1^N X_2^1$ is a positive combination of two different 1–body positive–semidefinite operators. Hence, $X^1 = \epsilon X_1^1 + X_2^1 \in \tilde{\mathcal{P}}_N^1$ and is not extreme which contradicts the assumption. Therefore, if $X^1$ is extreme, **ker** $\Gamma_1^N X^1$ is maximal.

To show the sufficiency, we assume that $\Gamma_1^N X^1 \geq 0$ has a maximal kernel but $X^1$ is not extreme in $\tilde{\mathcal{P}}_N^1$. Then, $X_1^1$ and $X_2^1 \in \tilde{\mathcal{P}}_N^1$ there exist such that $X^1 = \alpha X_1^1 + (1-\alpha) X_1^2$ , $0 < \alpha < 1$.Hence, $\Gamma_1^N X^1 = \alpha \Gamma_1^N X_1^1 + (1-\alpha)\Gamma_1^N X_2^1$ is a convex combination of two positive–semidefinite operators, and **ker** $\Gamma_1^N X^1 =$ **ker** $\Gamma_1^N X_1^1 \cap$ **ker** $\Gamma_1^N X_2^1$. Therefore, **ker** $\Gamma_1^N X^1 \subset$ **ker** $\Gamma_1^N X_1^1$ and **ker** $\Gamma_1^N X^1 \subset$ **ker** $\Gamma_1^N X_2^1$. If any of these inclusions is proper, then **ker** $\Gamma_1^N X^1$ is not maximal which contradicts the assumption.Thus, it remains to consider **ker** $\Gamma_1^N X^1 =$ **ker** $\Gamma_1^N X_1^1 =$ **ker** $\Gamma_1^N X_2^1$. Since the operators act on a finite dimensional Hilbert space, there exists $\epsilon > 0$ such that $\Gamma_1^N X^1 - \epsilon\, \Gamma_1^N X_1^1 \geq 0$ and **ker** $\Gamma_1^N X^1 \subset$ **ker** $\Gamma_1^N (X^1 - \epsilon X_1^1)$ which again contradicts the assumption that **ker** $\Gamma_1^N X^1$ is maximal. Hence $X^1$ is extreme. □

## Appendix 3

*type I collective behaviour*

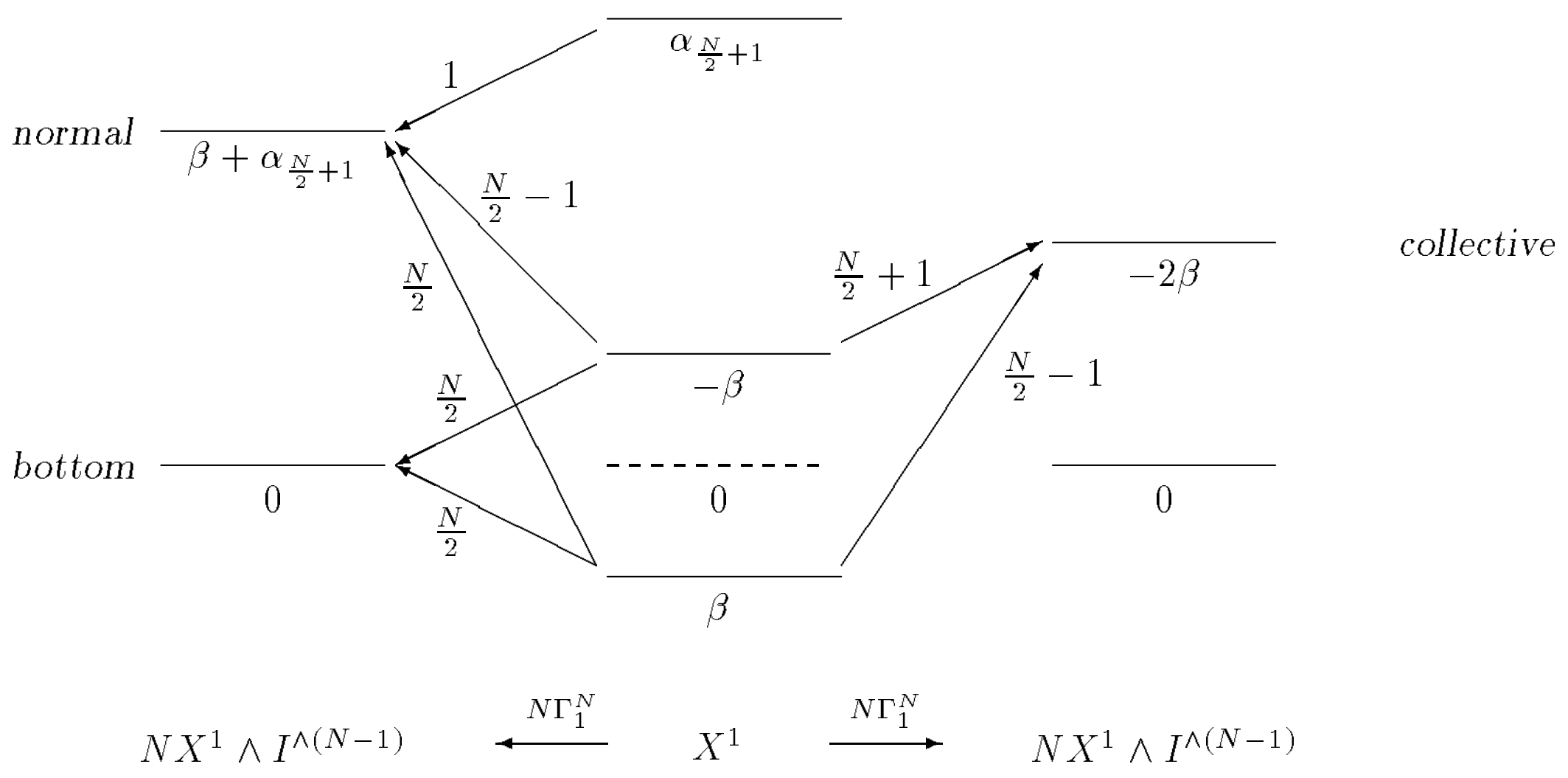

*type II collective behaviour*

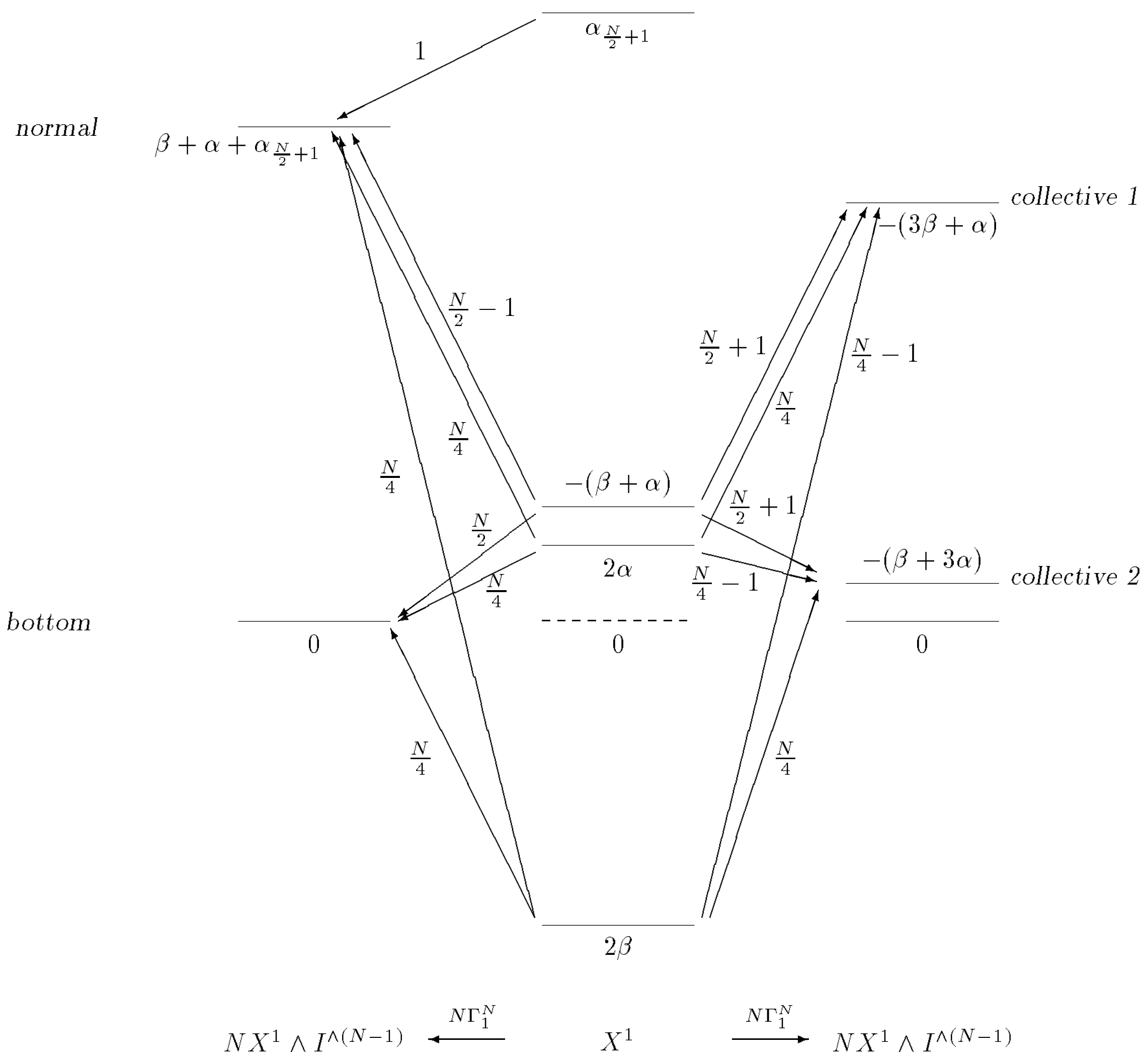

# References

[1] N.I. Akhiezer, I.M. Glazman, *Theory of Linear Operators in Hilbert Space*, Vol.I (1961), Vol.II (1963), Frederick Ungar Publishing Co., New York.

[2] P. Busch, P.J. Lahti, P. Mittelstaedt, *The Quantum Theory of Measurement*, Springer–Verlag, Berlin, Heidelberg, 1991.

[3] P. Busch, M. Grabowski, P. Lahti, Found. Phys. Lett. 2, 331 (1989)

[4] A.J. Coleman, Rev. Mod. Phys. 35, 668 (1963).

[5] A.J. Coleman, J. Math. Phys. 6, 1425 (1965).

[6] A.J. Coleman, J. Math. Phys. 13, 214 (1972).

[7] A.J. Coleman, Rep. Math. Phys. 4, 113 (1973).

[8] E.R. Davidson, J. Math. Phys. 10, 725 (1969).

[9] E.B. Davies, J.T. Lewis, Commun. math. Phys. 17, 239 (1970).

[10] E.B. Davies, *Quantum Theory of Open Systems*, Academic Press, 1976

[11] R.M. Erdahl, J. Math. Phys. 13, 1608 (1972).

[12] R.M. Erdahl, Int. J. Quant. Chem. 13, 697 (1978).

[13] R.M. Erdahl, H. Grudziński, Rep. Math. Phys. 14,405 (1978).

[14] R.M. Erdahl, in *Reduced Density Operators with Applications to Physical and Chemical Systems II*, R.M. Erdahl (Ed.), Queen's Papers on Pure and Applied Mathematics, No.40, Kingston, Ontario, 1974, p.13.

[15] R.M. Erdahl, V.H. Smith, Jr. (Eds), *Density Matrices and Density Functionals*, Proceedings of the A. John Coleman Symposium, D. Reidel Publishing Company, Dordrecht, Holland, 1987.

[16] C. Garrod, J.K. Percus, J. Math. Phys. 5, 1756 (1964).

[17] M. Grabowski, *Semi–spectral measures in non–relativistic quantum mechanics* (in Polish), Nicholas Copernicus Univ. Press, Toruń 1990

[18] H. Grudziński, Rep. Math. Phys. 8, 271 (1975).

[19] H. Grudziński, Rep. Math. Phys. 9, 199 (1976)

[20] H. Grudziński, Int. J. Quant. Chem. 27, 709 (1985).

[21] H. Grudziński, *A Study of the Convex Structure of the Set of Fermion N–representable 2–density Operators*, Nicholas Copernicus University Press, Torun, Poland, 1986 (p. 20).

[22] C. W. Helstrom, *Quantum Detection and Estimation Theory*, Academic Press, 1976

[23] A.S. Holevo, *Probabilistic and Statistical Aspects of Quantum Theory*, North Holland Publ. Co., Amsterdam, 1982.

[24] H.W. Kuhn, Proc. Sym. Appl. Math. 10, 141 (1960).

[25] H. Kummer, J. Math. Phys. 8, 2063 (1967).

[26] H. Kummer, Int. J. Quant. Chem. 12, 1033 (1977).

[27] H. Kummer, I. Absar, A.J. Coleman, J. Math. Phys. 18, 329, (1977).

[28] H. Kummer, I. Absar, J. Math. Phys. 18, 335 (1977).

[29] P-O. Löwdin, Phys. Rev. 97, 1474, 1490, 1512 (1955).

[30] G. Ludwig, *Foundation of Quantum Mechanics, Vol. I,II*, Springer Verlag, Berlin 1983, 1986.

[31] W.B. McRae, E.R. Davidson, J. Math. Phys. 13, 1527 (1972).

[32] R. Mc Weeny, Rev. Mod. Phys. 32, 335 (1960).

[33] M. Ozawa, J. Math. Phys. 25, 79 (1984).

[34] J. von Neumann, *Mathematical Foundations of Quantum Mechanics*, Princeton Univ. Press, Princeton, New Jersey, 1955.

[35] E. Prugovecki, *Quantum Mechanics in Hilbert Space*, Academic Press, New York, London, 1971.

[36] E. Prugovecki, *Stochastic Quantum Mechanics and Quantum Spacetime*, D. Reidel Publishing Company, Dordrecht 1986

[37] M.L. Yeseloff, H.W. Kuhn, J. Math. Phys. 10, 703 (1969).